\documentclass[]{aa}

\usepackage{hyperref}

\usepackage{natbib}

\usepackage{txfonts}
\usepackage{graphicx}

\begin{document}

\title{The bright end of the infrared luminosity functions and the abundance of hyperluminous infrared galaxies}

\author{L.~Wang$^{\ref{inst:SRON}, \ref{inst:Kapteyn}}$, F. Gao$^{\ref{inst:SRON}, \ref{inst:Kapteyn}}$,  P. N. Best$^{\ref{inst:SUPA}}$, K. Duncan$^{\ref{inst:SUPA}, \ref{inst:Leiden}}$, M. J. Hardcastle$^{\ref{inst:Hert}}$, R. Kondapally$^{\ref{inst:SUPA}}$, K. Ma{\l}ek$^{\ref{inst:NCNR}, \ref{inst:LAM}}$, I. McCheyne$^{\ref{inst:Sussex}}$, J. Sabater$^{\ref{inst:SUPA}}$, T. Shimwell$^{\ref{inst:Leiden}, \ref{inst:ASTRON}}$, C. Tasse$^{\ref{inst:GEPI}, \ref{inst:SA}, \ref{inst:USN}}$, M. Bonato$^{\ref{inst:Bologna}, \ref{inst:Padova}}$, M. Bondi$^{\ref{inst:Bologna}}$, R. K. Cochrane$^{\ref{inst:Harvard}}$,  D. Farrah$^{\ref{inst:Hawaii}, \ref{inst:IoA}}$, G. G\"urkan$^{\ref{inst:CSIRO}}$,  P. Haskell$^{\ref{inst:Hert}}$, W. J. Pearson$^{\ref{inst:SRON}, \ref{inst:Kapteyn}, \ref{inst:NCNR}}$, I. Prandoni$^{\ref{inst:Bologna}}$, H. J. A. R\"ottgering$^{\ref{inst:Leiden}}$, D. J. B. Smith$^{\ref{inst:Hert}}$, M. Vaccari$^{\ref{inst:Bologna}, \ref{inst:Cape}}$, W. L. Williams$^{\ref{inst:Leiden}}$}

\institute{\label{inst:SRON} SRON Netherlands Institute for Space Research, Landleven 12, 9747 AD, Groningen, The Netherlands \email{l.wang@sron.nl} 
\and \label{inst:Kapteyn} Kapteyn Astronomical Institute, University of Groningen, Postbus 800, 9700 AV Groningen, the Netherlands 
\and \label{inst:SUPA} SUPA, Institute for Astronomy, Royal Observatory, Blackford Hill, Edinburgh, EH9 3HJ, UK
\and \label{inst:Leiden} Leiden Observatory, Leiden University, PO Box 9513, 2300 RA Leiden, the Netherlands
\and \label{inst:Hert} Centre for Astrophysics Research, School of Physics, Astronomy and Mathematics, University of Hertfordshire, College Lane, Hatfield AL10 9AB, UK
\and \label{inst:NCNR} National Centre for Nuclear Research, ul. Pasteura 7, 02-093 Warszawa, Poland
\and \label{inst:LAM} Aix Marseille Univ. CNRS, CNES, LAM, Marseille, France
\and \label{inst:Sussex} Astronomy Centre, Dept. of Physics \& Astronomy, University of Sussex, Brighton BN1 9QH, UK
\and \label{inst:ASTRON} ASTRON, Netherlands Institute for Radio Astronomy, Oude Hoogeveensedijk 4, 7991 PD, Dwingeloo, The Netherlands
\and \label{inst:GEPI} GEPI, Observatoire de Paris, CNRS, Universit\'{e} Paris Diderot, 5 place Jules Janssen, 92190 Meudon, France
\and \label{inst:SA} Centre for Radio Astronomy Techniques and Technologies, Department of Physics and Electronics, Rhodes University, Grahamstown 6140, South Africa
\and \label{inst:USN} USN, Observatoire de Paris, CNRS, PSL, UO, Nan\c{c}ay, France
\and \label{inst:Bologna} INAF-Istituto di Radioastronomia, Via Gobetti 101, I-40129, Bologna, Italy
\and \label{inst:Padova} INAF-Osservatorio Astronomico di Padova, Vicolo dell'Osservatorio 5, I-35122, Padova, Italy.
\and \label{inst:Harvard} Harvard-Smithsonian Center for Astrophysics, 60 Garden St, Cambridge, MA 02138, USA
\and \label{inst:Hawaii} Department of Physics and Astronomy, University of Hawaii, 2505 Correa Road, Honolulu, HI 96822, USA
\and \label{inst:IoA} Institute for Astronomy, 2680 Woodlawn Drive, University of Hawaii, Honolulu, HI 96822, USA
\and \label{inst:CSIRO} CSIRO Astronomy and Space Science, PO Box 1130, Bentley WA 6102, Australia
\and \label{inst:Cape} Inter-University Institute for Data Intensive Astronomy, Department of Physics and Astronomy, University of the Western Cape, Robert Sobukwe Road, 7535 Bellville, Cape Town, South Africa
}

\date{Received / Accepted}

\abstract
   {}
   {We provide the most accurate estimate yet of the bright end of the infrared (IR) luminosity functions (LFs) and the abundance of hyperluminous IR galaxies (HLIRGs) with IR luminosities $>10^{13}L_{\odot}$, thanks to the combination of the high sensitivity, angular resolution, and large area of the LOFAR Deep Fields, which probes an unprecedented dynamic range of luminosity and volume.}
   {We cross-match  {\it Herschel} sources and LOFAR sources in Bo\"otes (8.63 deg$^2$), Lockman Hole (10.28 deg$^2$), and ELAIS-N1 (6.74 deg$^2$) with rms sensitivities of $\sim32$, 22, and 20 $\mu$Jy beam$^{-1}$, respectively. We divide the matched samples into 'unique' and 'multiple' categories. For the multiple matches, we de-blend the {\it Herschel} fluxes using the LOFAR positions and the 150-MHz flux densities as priors. We perform spectral energy distribution (SED) fitting, combined with multi-wavelength counterpart identifications and photometric redshift estimates,  to derive IR luminosities.}
   {The depth of the LOFAR data allows us to identify highly complete ($\sim92$\% completeness) samples of bright {\it Herschel} sources with a simple selection based on the 250 $\mu m$ flux density (45, 40, and 35 mJy in Bo\"otes, Lockman Hole, and ELAIS-N1, respectively). Most of the bright {\it Herschel} sources fall  into the unique category (i.e. a single LOFAR counterpart). For the multiple matches, there is  excellent correspondence between the radio emission and the far-IR emission. We find a good agreement in the IR LFs with a previous study out to $z\sim6$ which used de-blended {\it Herschel} data. Our sample gives the strongest and cleanest indication to date that the population of HLIRGs has surface densities of $\sim5$ to $\sim18$ / deg$^2$ (with variations due to a combination of the applied flux limit and cosmic variance) and an uncertainty of a factor of $\lesssim$2. In comparison, the GALFORM semi-analytic model significantly under-predicts the abundance of HLIRGs.}
   {}
   

\titlerunning{The bright end of the infrared luminosity function
}
\authorrunning{Wang et al.}

\maketitle

\section{Introduction}

A key discovery from the Infrared Astronomical Satellite (IRAS; Neugebauer et al. 1984) was the existence of dusty galaxies with remarkably high infrared (IR) luminosity $L_{\rm IR}$ (integrated between 8 and 1000 $\mu m$).  Ultraluminous IR galaxies (ULIRGs) are defined as galaxies with $L_{\rm IR}>10^{12}L_{\odot}$ and star-formation rates (SFRs) around hundreds of solar masses per year. In the nearby Universe, ULIRGs, which are among the brightest objects, are mostly interacting systems (e.g. Armus, Heckman \& Miley 1987; Melnick \& Mirabel 1990; Hutchings \& Neff 1991; Clements et al. 1996; Murphy et al. 1996; Surace, Sanders \& Evans 2000; Farrah et al. 2001; Veilleux, Kim \& Sanders 2002). The power source in these galaxies is usually a combination of dusty star formation and mass accretion onto the central supermassive black hole (e.g. Rowan-Robinson \& Crawford 1989; Smith, Lonsdale \& Lonsdale 1998; Klaas et al. 2001; Farrah et al. 2003, 2007; Franceschini et al. 2003; Ptak et al. 2003; Imanishi et al. 2007).  Even more extreme than ULIRGs are hyperluminous IR galaxies (HLIRGs), galaxies with $L_{\rm IR}>10^{13}L_{\odot}$ and SFRs around thousands of solar masses per year (Cutri et al. 1994; Frayer et al. 1999; Rowan-Robinson 2000; Farrah et al. 2002; Wilman et al. 2003; Rowan-Robinson \& Wang 2010). Many HLIRGs show evidence of active galactic nucleus (AGN) dust torus emission, but the far-IR (FIR) and sub-millimetre (sub-mm) emission is usually driven by star formation (Farrah et al. 2002; Verma et al. 2002; Rowan-Robinson \& Wang 2010; Ruiz et al. 2010; Efstathiou et al. 2014).

Since the launch of IRAS, a number of IR and sub-mm surveys using ground- and space-based facilities -- such as the Submillimeter Common-User Bolometer Array (SCUBA) at the James Clerk Maxwell Telescope (JCMT), the {\it Planck} satellite, the {\it Herschel} Space Telescope, the Atacama Cosmology Telescope (ACT), and the South Pole Telescope (SPT) -- have been effective in detecting large numbers of highly luminous IR galaxies with $L_{\rm IR}>10^{12-14} L_{\odot}$ over a wide range of redshifts (e.g. Bussman et al. 2013; Wardlow et al. 2013; Weiss et al. 2013; Vieira et al. 2013; Planck Collaboration XXVII 2015; Canameras et al. 2015; Planck Collaboration XXXIX 2016; Rowan-Robinson et al. 2016, 2018).  It has been found that ULIRGs and HLIRGs are much more common at high redshifts  than in the local Universe (Chapman et al. 2003, 2005; Wardlow et al. 2011; Casey et al. 2012; Magnelli et al. 2013) and are likely to be the progenitors of today's massive elliptical galaxies (Micha{\l}owski, Hjorth \& Watson 2010; Hickox et al. 2012; Wang et al. 2013).

Infrared luminous galaxies provide stringent tests for theoretical models of galaxy assembly (and are particularly constraining for feedback models). The issue for the models is that neither mergers nor cold accretion could produce SFRs in excess of a few thousand solar masses per year -- mergers because they cannot channel enough gas to the centres of haloes (Narayanan et al. 2010; Dav{\'e}  et al. 2010), and cold accretion because massive haloes inhibit the gas flow onto central galaxies via shock heating (Kere{\v{s}} et al. 2005; Narayanan et al. 2015). The models could potentially reproduce the highest SFRs observed if feedback is turned off completely, but they would then strongly over-predict the local galaxy stellar mass function.

The main goal of this paper is to identify robust and complete samples of IR luminous galaxies over large areas to provide the most accurate determination yet of the abundance of HLIRGs and the bright end of the IR luminosity function (LF) by exploiting the combined power of deep {\it Herschel} extragalactic surveys and the Low Frequency Array (LOFAR; R{\"o}ttgering et al. 2011; van Haarlem et al. 2013) Two-metre Sky Survey (LoTSS) Deep Fields (Tasse et al 2020; Sabater et al 2020; Kondapally et al 2020). From {\it Herschel} observations, one can select large samples of IR luminous sources close to the peak of the IR spectral energy distribution (SED) and which are therefore minimally affected by selection bias caused by variations of the SED shapes. However, due to the large beam\footnote{The full width at half maximum (FWHM) of the beams of the Spectral and Photometric Imaging Receiver (SPIRE) aboard {\it Herschel} are 18.1\arcsec, 25.2\arcsec , and 36.6\arcsec\, at 250, 350, and 500 $\mu$m, respectively (Swinyard et al. 2010).}, it is very challenging to reliably match {\it Herschel} sources with galaxies detected in much higher resolution deep optical and near-IR (NIR) data. Moreover, the {\it Herschel} fluxes of blindly detected sources could be boosted due to blending issues, which means multiple sources can blend together inside the beam. Consequently, the intrinsically fainter sources could artificially boost the bright end of the IR LF (e.g. Koprowski et al. 2017; Gruppioni \& Pozzi 2019). Many studies have addressed these issues using various techniques, such as the stacking method (e.g. B{\'e}thermin et al. 2010; Wang et al. 2016a), the P(D) measurements (e.g. Patanchon et al. 2009; Glenn et al. 2010), or source extraction using prior information extracted from surveys with higher angular resolution (e.g. Roseboom et al. 2010; Wang et al. 2013, 2014). However, there are important limitations to these methods. For example, statistical methods lose information on individual galaxies, while prior-based methods rely on assumptions of the SEDs (Pearson et al. 2017, 2018; Liu et al. 2018; Wang et al. 2019a). 

In this paper, we use deep LOFAR data at 150 MHz in three {\it Herschel} survey fields  to identify robust samples of IR luminous galaxies, with a simple 250 $\mu m$ selection and a high level of completeness.  It is well known that high-resolution 1.4 GHz  radio data led to a breakthrough in the multi-wavelength counterpart identification of the sub-mm sources identified in single-dish surveys with large beam-size (Ivison et al. 2002) thanks to the FIR to radio correlation (FIRC), which shows little evidence of (strong) redshift evolution (e.g. Jarvis et al. 2010; Sargent et al. 2010; Bourne et al. 2011; Magnelli et al. 2015; Wang et al. 2019b). In a similar vein, thanks to the much higher resolution of LOFAR compared to the {\it Herschel} beam and the depths of the LoTSS Deep Fields,  we can more easily de-blend {\it Herschel} sources, which are significantly affected by multiplicity. Moreover, the precise positions of the LOFAR sources allow us to unambiguously match {\it Herschel} sources with their optical and NIR counterparts and to derive physical properties such as $L_{\rm IR}$. 

The rest of the paper is organised as follows. In Sect. 2, we describe the FIR and radio catalogues as well as the value-added products from the LOFAR and {\it Herschel} surveys used in this study. In Sect. 3, we explain the cross-matching process of the {\it Herschel} and LOFAR catalogues, the division of {\it Herschel}-LOFAR matches into 'unique' and 'multiple' categories, and the de-blending of {\it Herschel} fluxes for the multiple matches. We also discuss the impact of multiplicity and gravitational lensing. In Sect. 4, we present our measurement of the bright end of the IR LFs and comparisons with previous measurements. In Sect. 5, we show our results on the abundance and redshift distribution of HLIRGs and discuss the discrepancy between the observations and the predictions from the GALFORM semi-analytic models. Finally, we give our conclusions in Sect. 6. We assume Wilkinson Microwave Anisotropy Probe year 7 (WMAP7) cosmology with $\Omega_m=0.272$,  $\Omega_{\Lambda}=0.728$,  and $H_0=70.4$ km s$^{-1}$ Mpc$^{-1}$ (Komatsu et al. 2011) to be consistent with the SED modelling and fitting tool used in this paper.

\section{Data}

\subsection{The LoTSS Deep Fields and the associated multi-wavelength datasets}

LoTSS is an ongoing sensitive, high-resolution, low-frequency (120-168 MHz) radio continuum imaging survey of the northern sky (Shimwell et al. 2017, 2019). It provides the astrometric precision needed for the accurate and robust identification of optical and NIR counterparts (e.g. McAlpine et al. 2012) and a sensitivity superior to previous wide area surveys at higher frequencies, such as the NRAO Very Large Array Sky Survey (NVSS; Condon et al. 1998) and Faint Images of the Radio Sky at Twenty Centimeters (FIRST) survey (Becker, White, \& Helfand 1995). The primary observational objectives are to reach a sensitivity of $<100\mu$Jy/beam at an angular resolution, defined as the full width at half maximum (FWHM) of the synthesised beam, of $\sim6\arcsec$ at optimal declinations. In addition, LoTSS repeatedly observes several northern fields that have the best multi-wavelength data coverage over several-degree scales. Ultimately, these LoTSS Deep Fields will reach depths of $\sim10\mu$Jy/beam over a combined area of many tens of square degrees.

In this paper, we make use of the first LoTSS Deep Fields data release (Tasse et al 2020; Sabater et al 2020), which consists of deep imaging and source catalogues at 150 MHz down to an rms sensitivity of $\sim$20, 22, and 32 $\mu$Jy per beam over three regions: the European Large Area ISO Survey-North 1 (ELAIS-N1; Oliver et al. 2000), Lockman Hole (LH), and Bo\"otes (Jannuzi \& Dey 1999). The depth of the LOFAR data is comparable to the deepest existing radio continuum surveys (e.g. VLA-COSMOS over 2 deg$^2$; Smol{\v{c}}i{\'c} et al. 2017) but with more than an order of magnitude larger areal coverage. Therefore, the LoTSS Deep Fields offer us an unprecedented opportunity to study AGNs and star-forming galaxies in statistically large samples covering a large dynamic range in radio luminosity out to the highest redshifts.

As the three fields benefit from extensive panchromatic data, the first LoTSS Deep Fields data release also includes cross-matched optical and IR catalogues in the overlapping multi-wavelength regions as well as value-added products such as photometric redshifts. Kondapally et al. (2020) present new optical to mid-IR photometry catalogues and robust cross-identification with the LOFAR radio source population. The new, forced, matched-aperture multi-wavelength catalogues in ELAIS-N1 and LH were generated by first re-sampling available optical-IR imaging data into a common pixel scale and generating a mosaic per filter using \textsc{SWarp}. Source detection was then performed on deep $\chi^{2}$ detection images (also built using \textsc{SWarp}), which incorporates flux information from the available imaging datasets, with forced photometry performed at these positions across all filters, to generate catalogues. In Bo\"otes, Kondapally et al. (2020) made use of the existing point spread function (PSF) matched I-band and 4.5-$\mu$m band detected catalogues from Brown et al. (2007, 2008) to generate a similar multi-wavelength catalogue for the radio cross-matching. The radio-optical cross-matching was performed using a combination of the statistical likelihood ratio method (de Ruiter et al. 1977; Sutherland \& Saunders 1992) and visual classification, finding counterparts for $>$ 97\% of the radio-detected sources. Duncan et al. (2020) present consistent photometric redshift estimates for the optical source catalogues, using a hybrid methodology optimised to produce the best possible performance for the radio continuum-selected sources and the wider optical source population. Compared to spectroscopic redshifts, the photometric redshifts show a smaller scatter of 1.6 - 2\% for galaxies and 6.4 - 7\% for AGNs as well as an outlier fraction (defined as $\delta z/(1+z)>0.15$) of 1.5-1.8\% for galaxies and 18 - 22\% for AGNs (Duncan et al. 2020).  In general, the photo-$z$ scatter and outlier fraction deteriorate with increasing redshift for the host-dominated population. For the AGN population, the photo-$z$ scatter and outlier fraction do not evolve significantly with redshift, and the photo-$z$ estimates are expected to be accurate out to $z>6$. In this paper, we used \textsc{flag\_clean} $\ne$3 to exclude sources in the Spitzer star mask region, in conjunction with \textsc{flag\_overlap} to select sources with reliable photometry in the majority of the multi-wavelength bands\footnote{\textsc{flag\_clean}  is the bright star masking flag indicating masked and unmasked regions in the {\it Spitzer}- and optical-based bright star mask. \textsc{flag\_overlap}  is the overlap bit flag in the optical to mid-IR photometry catalogues and the radio cross-match catalogues indicating the multi-wavelength coverage of a given source. Further details regarding these flags are available in Kondapally et al. (2020).}. The resulting selected areas are 6.74 deg$^2$, 10.28 deg$^2$, and 8.63 deg$^2$ for ELAIS-N1, LH, and Bo\"otes, respectively (Kondapally et al. 2020). The cross-identification rates  are 97.6\%, 97.6\%, and 96.9\% over the selected areas in ELAIS-N1, LH, and Bo\"otes, respectively (Kondapally et al. 2020).

\subsection{The {\it Herschel} surveys of the LoTSS Deep Fields}

The FIR and sub-mm data in the LoTSS Deep Fields were taken from the {\it Herschel} Multi-tiered Extragalactic Survey (HerMES; Oliver et al. 2012), which is the largest guaranteed time key programme on the {\it Herschel} Space Observatory (Pilbratt et al. 2010). HerMES has a wedding cake survey design that consists of known clusters and nested blank fields (ranging from 0.01 to $\sim$ 20 deg$^2$) observed with the Spectral and Photometric Imaging Receiver (SPIRE; Griffin et al. 2010) at 250, 350, and 500 $\mu m$ and with the Photodetector Array Camera and Spectrometer (PACS; Poglitsch et al. 2010) at 100 and 160 $\mu m$ for a subset of the HerMES fields.  The three LoTSS Deep Fields are part of the Level 5 and 6 tier observations of HerMES. The 5-$\sigma$ instrumental noise is around  14, 11, and 16 mJy at 250, 350, and 500 $\mu m$, respectively, for Level 5, and around 26, 21, and 30 mJy at 250, 350, and 500 $\mu m$, respectively, for Level 6 (Oliver et al. 2012; Wang et al. 2014). The 5$\sigma$ confusion noise is 24.0, 27.5, and 30.5 mJy at 250, 350, and 500 $\mu m$, respectively (Nguyen et al. 2010), which is approximately equivalent to the depth of Level 6 observations. Therefore, the 5-$\sigma$ total noise (by adding instrumental noise and confusion noise in quadrature) is around 28 and 35 mJy at 250 $\mu m$ for Level 5 and 6 fields, respectively. As will be presented in Sect. 3, our ${\it Herschel}$-selected samples matched to LOFAR sources are at or above 5-$\sigma$ in total noise.

{\it Herschel}-selected sources represent excellent samples of dusty star-forming galaxies as the {\it Herschel} bands cover the peak of the thermal SED from interstellar dust over a wide redshift range. Therefore, {\it Herschel}-selected sources are perfect for performing an accurate and complete census of the underlying IR-luminous population. Sub-mm galaxies selected at 850 $\mu m$ suffer from selection bias against SED types characterised by warmer dust temperatures, known as the sub-mm dust-temperature selection effect (Blain 1996; Eales et al. 2000; Blain et al. 2004). In contrast, the {\it Herschel} wavebands (especially the 250 $\mu m$ band) are minimally affected by the variations of SED shapes. For example, for a galaxy with $L_{\rm IR}=10^{12.5} L_{\odot}$ at $z=2$, the observed 250 $\mu m$ flux density only varies by a factor of $\lesssim2$ for SED shapes characterised by effective dust temperatures ranging from $\sim15$ to 80 $K$ (Casey, Narayanan \& Cooray 2014), which spans the range of observed temperatures (e.g. Symeonidis et al. 2013; Schreiber et al. 2018; Zavala et al. 2018). In comparison, the 850 $\mu m$ flux density changes by a factor of $\sim30$ over the same temperature range. Of course, selecting by a given 250 $\mu m$ flux threshold will lead to an increasing IR luminosity threshold towards higher redshift.

Driven by the requirement of a simple selection with which to understand completeness levels, we made use of the blind catalogues (DMU22) from the {\it Herschel} Extragalactic Legacy Project (HELP\footnote{http://hedam.lam.fr/HELP/}, Shirley et al. 2019, Oliver et al., in prep.) database, which collates and homogenises observations from many astronomical observatories to provide an integrated dataset covering a wide range of wavelengths, with its key focus on the data from {\it Herschel} extragalactic surveys. Briefly, the blind sources are selected by finding peaks on the matched filtered (MF) maps (Chapin et al 2011). Only peaks with flux densities above the 85\% completeness level at every SPIRE wavelength individually are saved.  The {\it Herschel} data have similar depths across the LoTSS Deep Fields. To determine accurate centres for each source in the blind catalogue, the Pearson correlation coefficient on sub-pixel positions around the original detection is calculated using all three SPIRE bands. The best-fit flux density for a given source at the updated position is then determined using inverse variance weighting.  Finally, the three SPIRE catalogues are combined by removing duplicates at 350 and 500 $\mu$m using a nearest neighbour matching algorithm with 12\arcsec\, and 18\arcsec\, radii, respectively, adopting the position in the shortest wavelength available for each merged source.

\section{The {\it Herschel}-LOFAR cross-matched samples}

First, we cross-matched the blind {\it Herschel}-SPIRE catalogues with LOFAR sources using a cross-matching radius of 6\arcsec. The 6\arcsec\, radius results in low false identification rates of 5.4\%, 4.4\%, and 4.2\% in ELAIS-N1, Bo\"otes, and LH, respectively. The false identification rate is derived by comparing the number of {\it Herschel}-LOFAR matches with the number of matches between a randomised {\it Herschel} catalogue, by perturbing the positions by up to 30\arcmin  \ in right ascension (RA) and declination (DEC), and the LOFAR catalogue. 

Then we checked the fraction of LOFAR-matched {\it Herschel} sources as a function of the 250 $\mu$m flux density,  $S_{250\mu m}$. In ELAIS-N1, 92.2\% of the sources with $S_{250\mu m}> 35$ mJy in the {\it Herschel} blind catalogue are matched to LOFAR sources. In LH, 92.3\% of the sources with $S_{250\mu m}> 40$ mJy are matched to LOFAR sources. In Bo\"otes, 91.3\% of the sources with  $S_{250\mu m}> 45$ mJy are matched to LOFAR sources. Therefore, we adopted 250 $\mu m$ flux density limits of 35, 40, and 45 mJy in ELAIS-N1, LH, and Bo\"otes, respectively. The ELAIS-N1 field is the deepest, followed by LH and Bo\"otes. This is in agreement with the depths of the LOFAR data in these fields. At our adopted flux density limit, our  {\it Herschel}-LOFAR cross-matched samples are highly reliable (characterised by a low false identification rate $\sim5\%$) and complete (>90\%), with a simple selection based on a flux limit at 250 $\mu$m.

\subsection{'Unique' {\it Herschel}-LOFAR matches}

\begin{figure}
\includegraphics[height=2.3in,width=3.3in]{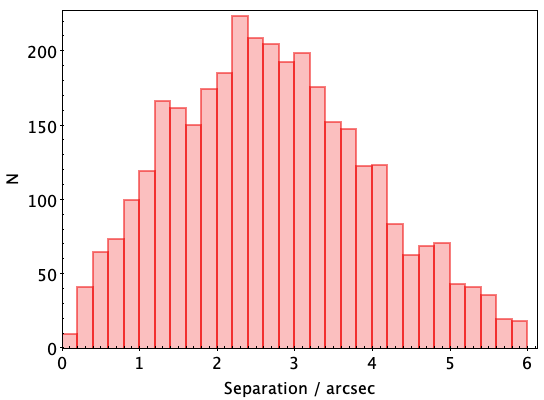}
\caption{Distribution in positional separation of the unique {\it Herschel}-LOFAR matches in ELAIS-N1. The other two fields show similar distributions.}
\label{sep}
\end{figure}

\begin{figure}
\includegraphics[height=2.5in,width=3.4in]{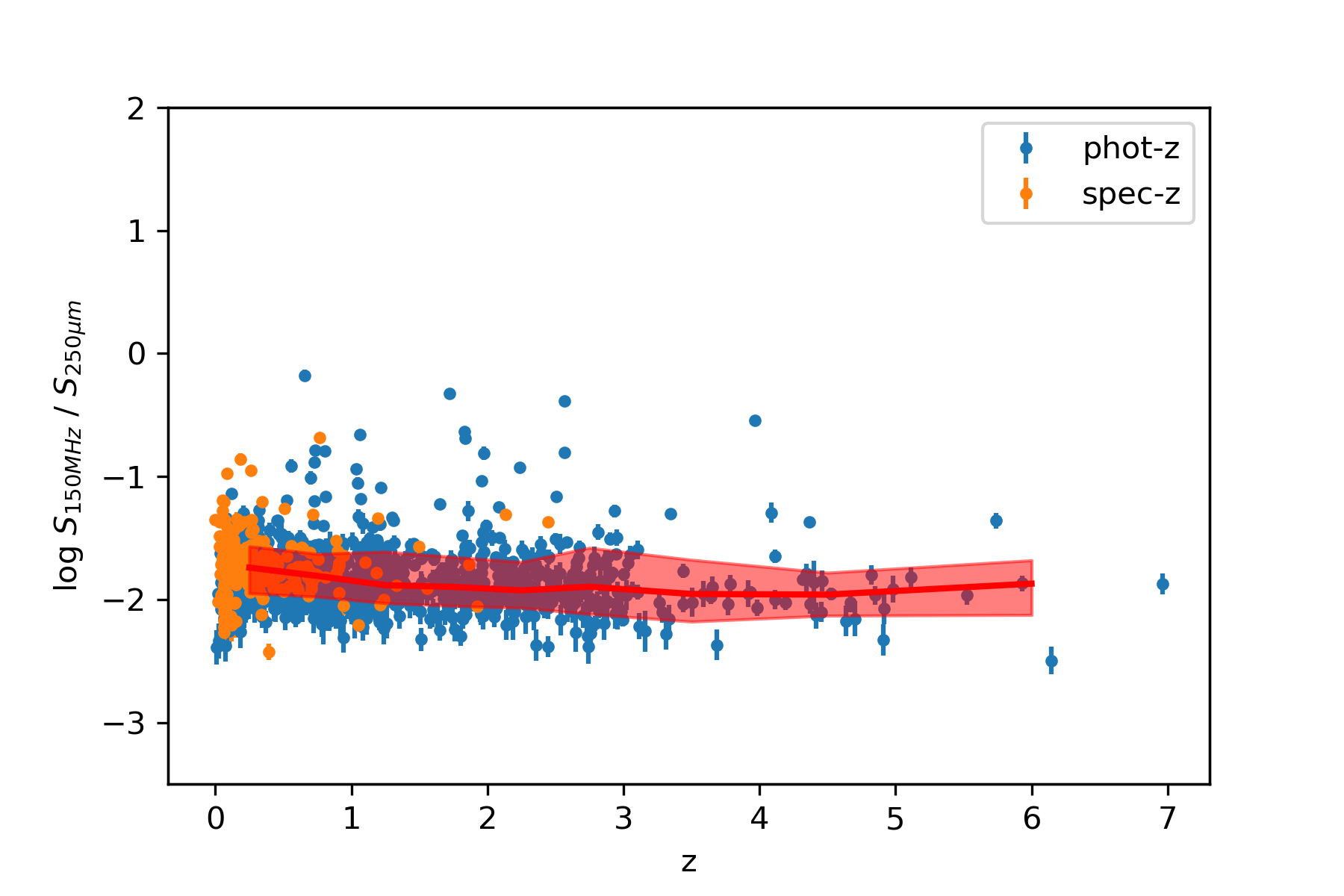}
\includegraphics[height=2.5in,width=3.4in]{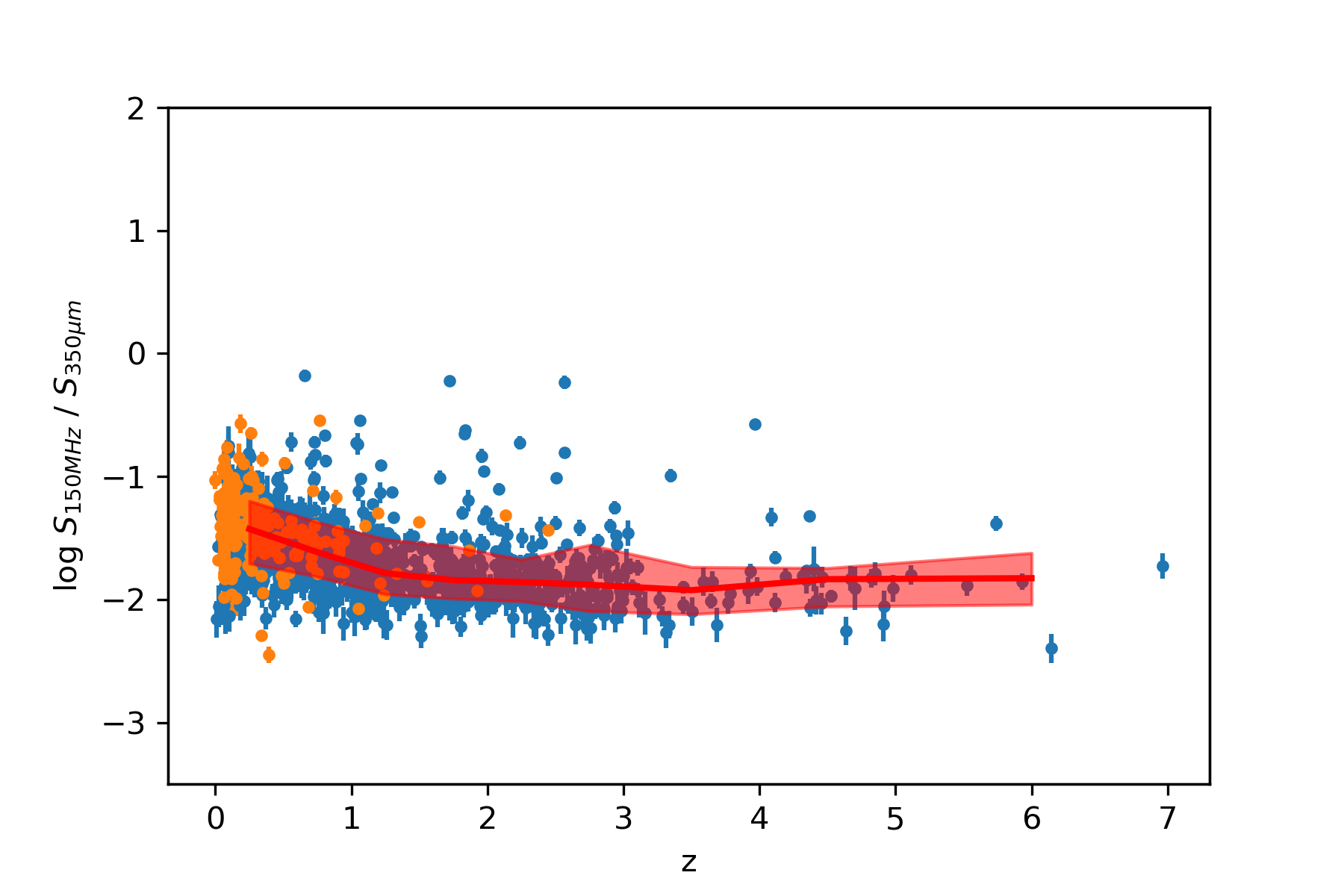}
\includegraphics[height=2.5in,width=3.4in]{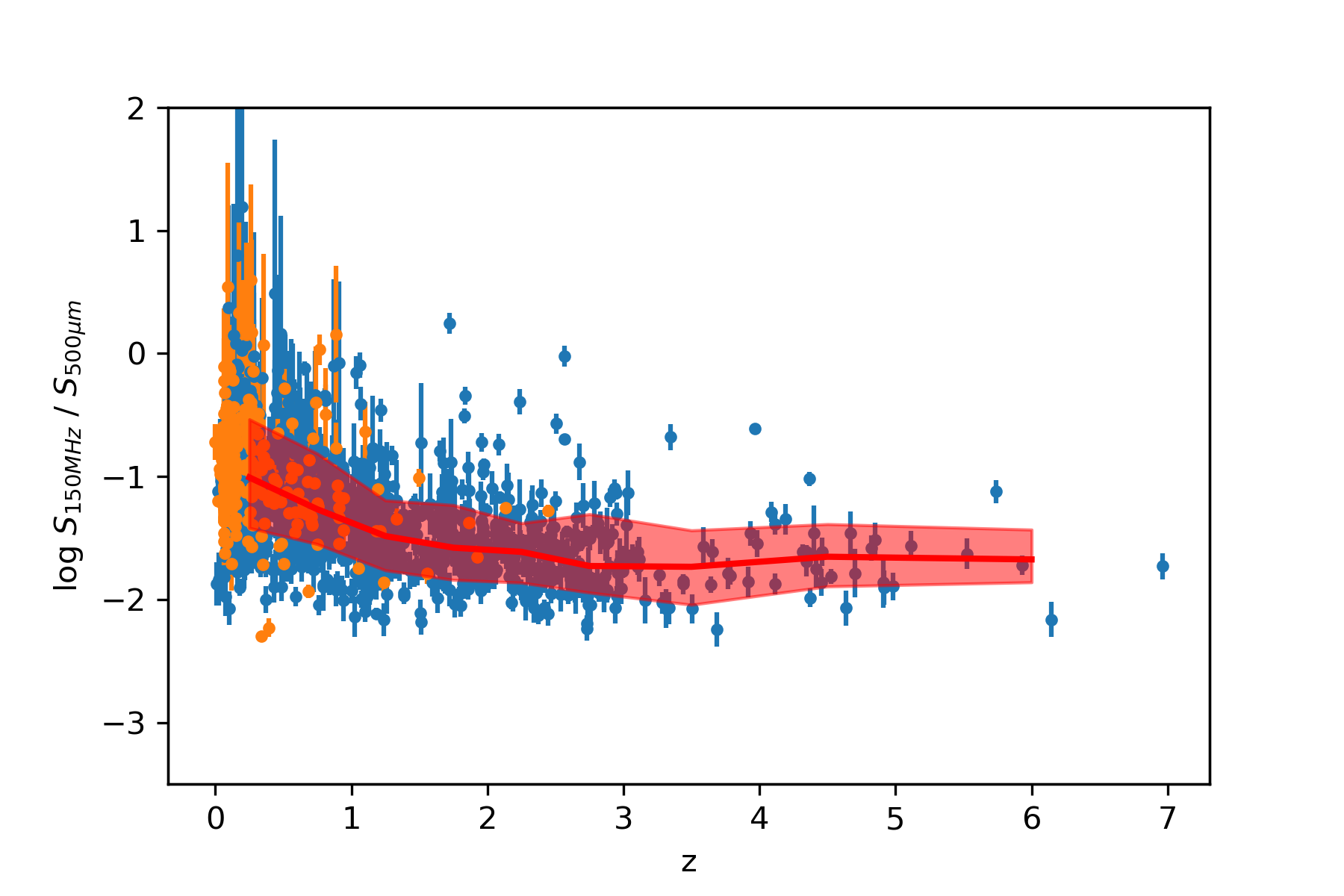}
\caption{Flux ratios between the 150 MHz flux density and the flux densities at 250, 350, and 500 $\mu$m as a function of redshift (colour-coded by photometric or spectroscopic redshift) in ELAIS-N1. The red line shows the median flux ratios in bins of redshift (with the redshift bin width set to 0.5 at $z<3$ and 1 at $z>3$), and the shaded band shows the range between the 16th and 84th percentiles. The other two fields show similar patterns.}
\label{fratio}
\end{figure}

We then separated the {\it Herschel}-LOFAR matches into unique and multiple categories. The subset of unique matches is composed of {\it Herschel} sources with a single LOFAR counterpart within the {\it Herschel} beam of 18\arcsec\, at 250 $\mu$m.  Figure~\ref{sep} shows the distribution in positional separations of the unique matches in ELAIS-N1, which peaks at $\sim2.2\arcsec$. As the LOFAR positional errors are negligible in comparison, this is consistent with the observed positional error of ${\it Herschel}$ 250 $\mu m$-selected sources (Smith et al. 2011; Bourne et al. 2016) and the theoretical expectation that the positional error depends on the FWHM and the signal-to-noise ratio (S/N) of the detected sources (Ivison et al. 2007),
\begin{equation}
\sigma_{\rm pos} ({\rm S/N}) = 0.6 \frac{{\rm FWHM}}{{\rm S/N}},
\end{equation}
where {\rm the S/N} is close to 5 for most of our sources (nearly half with $5 < {\rm S/N} < 6-7$).

For the unique matches, we simply used the SPIRE fluxes provided in the DMU22 blind catalogues as these matches are expected to contain bright {\it Herschel} sources that are least affected by issues of multiplicity. Figure~\ref{fratio} shows the flux ratios of the unique matches between the 150-MHz flux density and the flux densities at 250, 350, and 500 $\mu$m as a function of redshift in ELAIS-N1. For the majority of the unique matches, there is a clear correlation between the flux densities at 250, 350, and 500 $\mu$m and the 150-MHz flux density over the plotted redshift range, although matches with spectroscopic redshifts are mostly limited to $z<1$. This is expected from the FIRC and further supports the reliability of our samples  cross-matched between {\it Herschel} and LOFAR.

\subsection{'Multiple' {\it Herschel}-LOFAR matches}

The subset of multiple matches is composed of {\it Herschel} sources with more than one LOFAR counterpart within the {\it Herschel} beam. For the {\it Herschel} sources in the multiple matches category, we could no longer use the raw flux from the DMU22 blind catalogues as these sources may be significantly affected by multiplicity. Thanks to the much higher angular resolution of the LOFAR data, we can now see the individual galaxies that make up a single {\it Herschel} detection. We note that it is extremely difficult, or impossible, to identify all the contributing galaxies (regardless of their flux density) associated with the same detection. However, we are only interested in identifying the contributing galaxies that are above our flux limit. Based on the depths of the LoTSS Deep Fields and the correlation between the 150-MHz flux density and the 250-$\mu m$ flux density in Fig.~\ref{fratio}, we expect that our LOFAR observations can detect multiples with flux ratios as high as 3 - 4 for the faintest 250 $\mu m$ sources in our samples, which means the $L_{\rm IR}$ of the galaxies in our samples could be revised down by at most 20 - 25\% (as there could be further multiples with flux ratios $>3$ - 4).  

To assign the correct amount of {\it Herschel} flux to the LOFAR sources associated with the same {\it Herschel} source, we ran our SED prior enhanced XID+, which is a probabilistic de-blending tool (Hurley et al. 2017; Pearson et al. 2017, 2018), in regions around the LOFAR sources on the SPIRE maps. By adding informative (but not overly restrictive) priors on the IR SED based on the associated multi-wavelength data, Pearson et al. (2017) have demonstrated (using ALMA 870 $\mu$m continuum data as the ground truth) that the SED prior-enhanced de-blending method outperforms the standard form of XID+, which uses a flat flux prior between zero and the brightest value on the map. In this paper, we take a different approach in applying priors on the IR SED in order to exploit the FIRC. We incorporated the observed 150-MHz flux densities in the SED priors to improve the de-blending of {\it Herschel} fluxes. Specifically, we made use of the best fit as a function of redshift and two times the standard deviations in the flux ratios, as plotted in Fig.~\ref{fratio}, to form estimates of the flux densities and uncertainties at 250, 350, and 500 $\mu$m. Combined with the precise positional information of the LOFAR sources, XID+ then uses the Bayesian inference tool Stan (Stan Development Team 2015a,b) to obtain the full posterior probability distribution on flux density estimates by modelling the confusion-limited SPIRE maps. In the appendix, we show example {\it Herschel} and LOFAR images of the multiple matches with the highest level of multiplicity to demonstrate the effectiveness of our approach. 

\begin{figure}
\includegraphics[height=2.5in,width=3.4in]{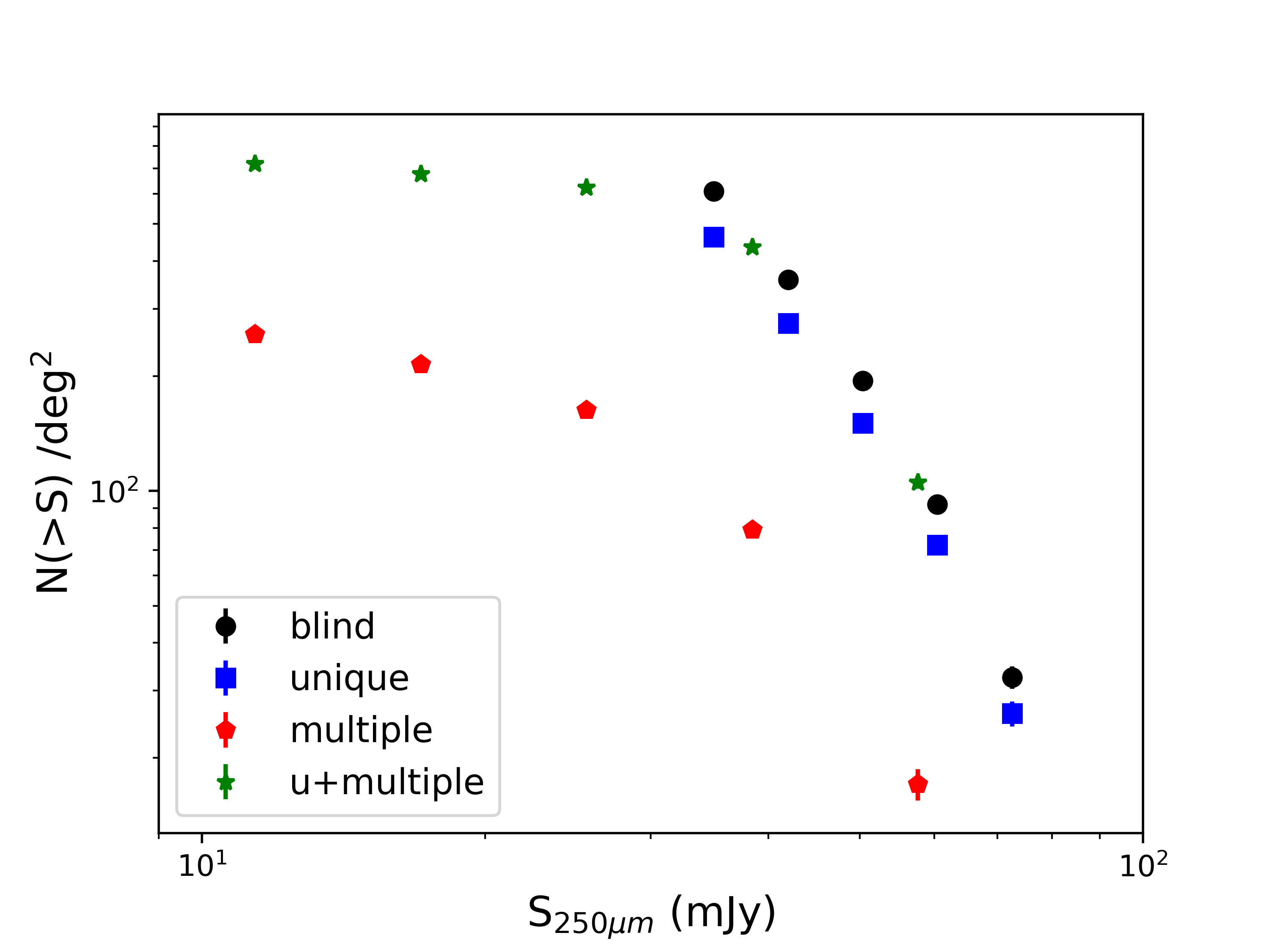}
\includegraphics[height=2.5in,width=3.4in]{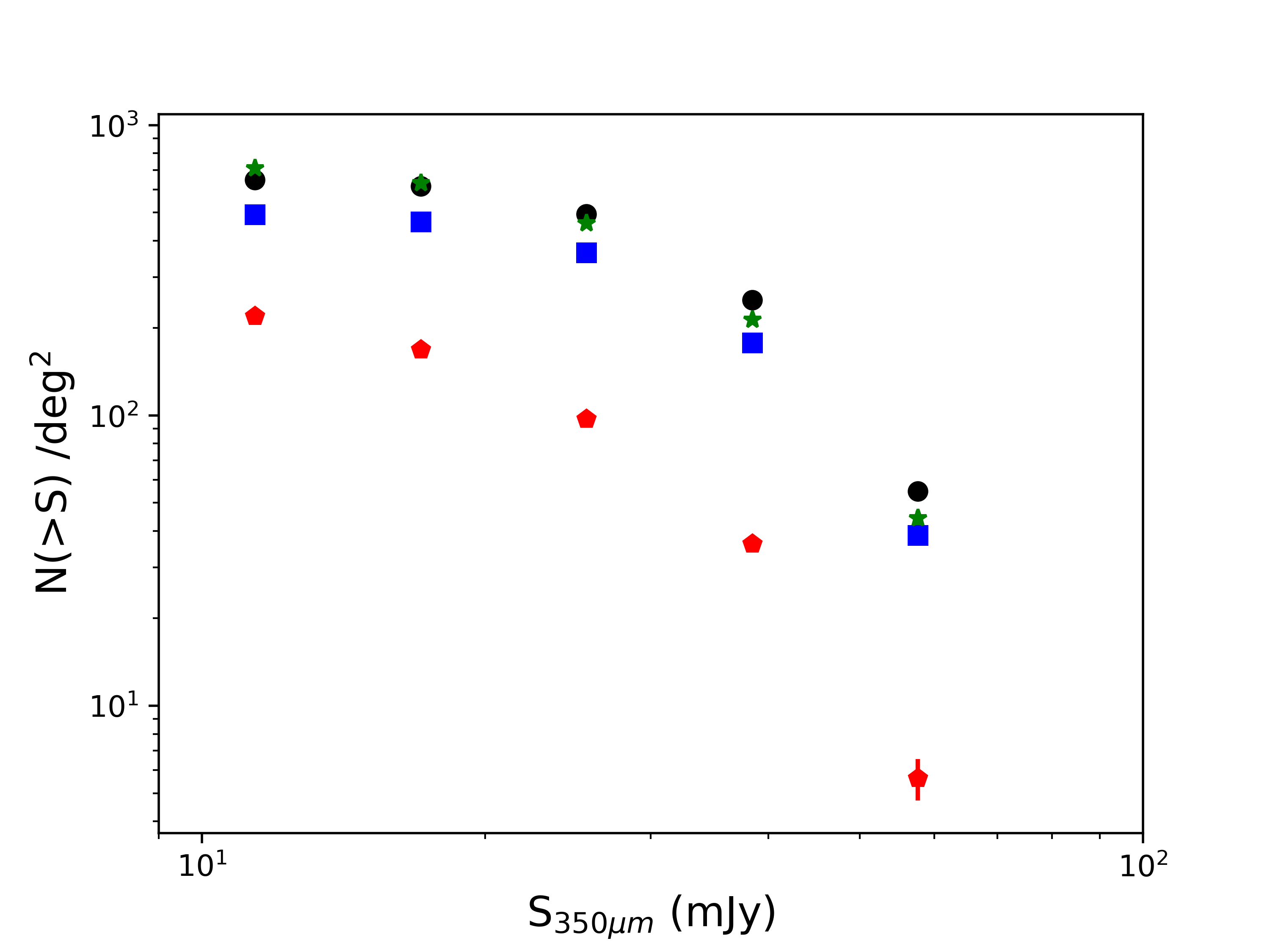}
\includegraphics[height=2.5in,width=3.4in]{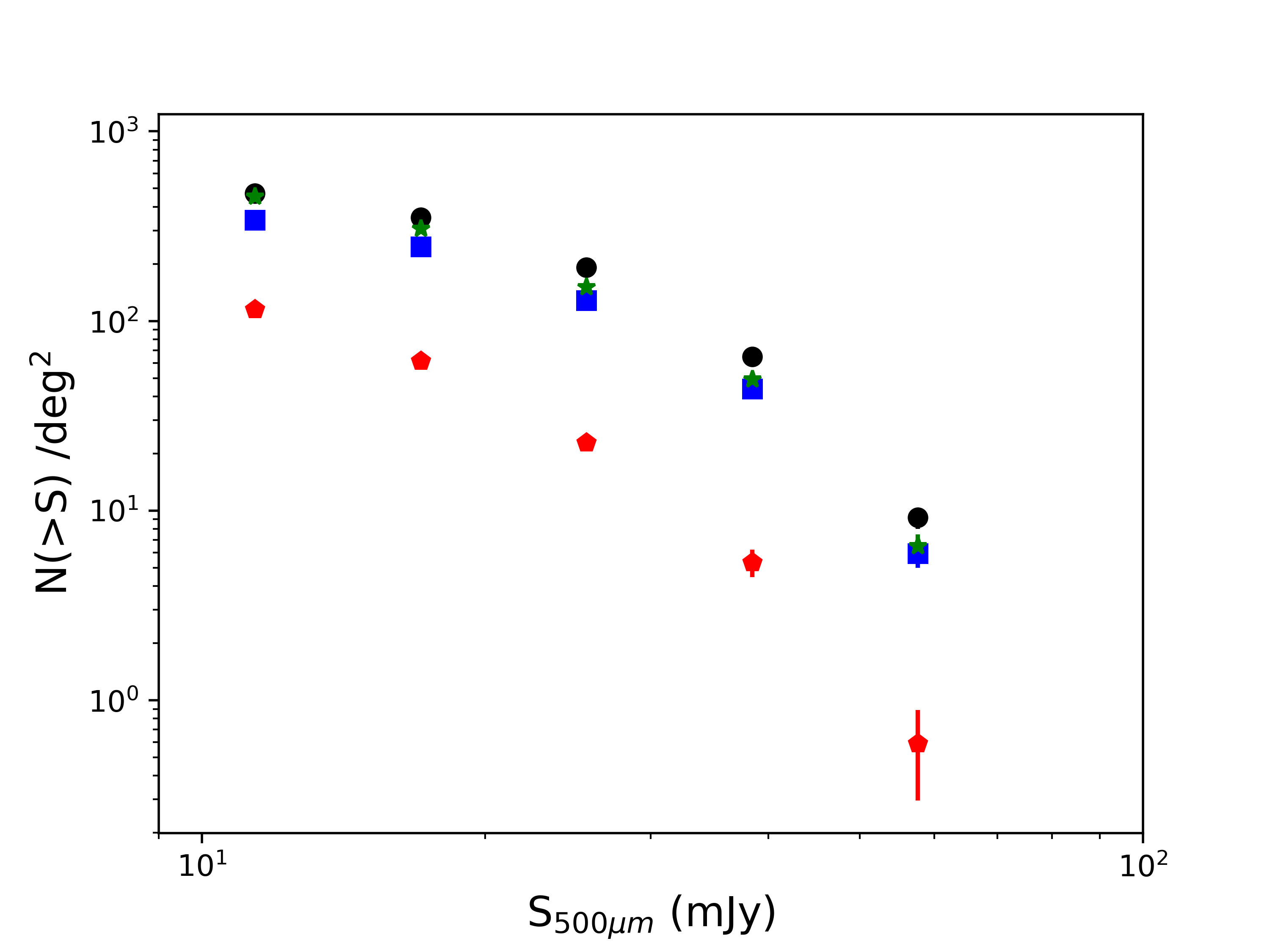}
\caption{Integral number counts at 250, 350, and 500 $\mu$m (from top to bottom) in ELAIS-N1. The other two fields show similar number counts. The black symbols show the raw number counts from the {\it Herschel} catalogues. The unique (multiple) {\it Herschel}-LOFAR matches are shown in blue (red). The total number counts of the unique and multiple matches are shown in green.}
\label{int_counts}
\end{figure}

\begin{figure}
\includegraphics[height=2.5in,width=3.4in]{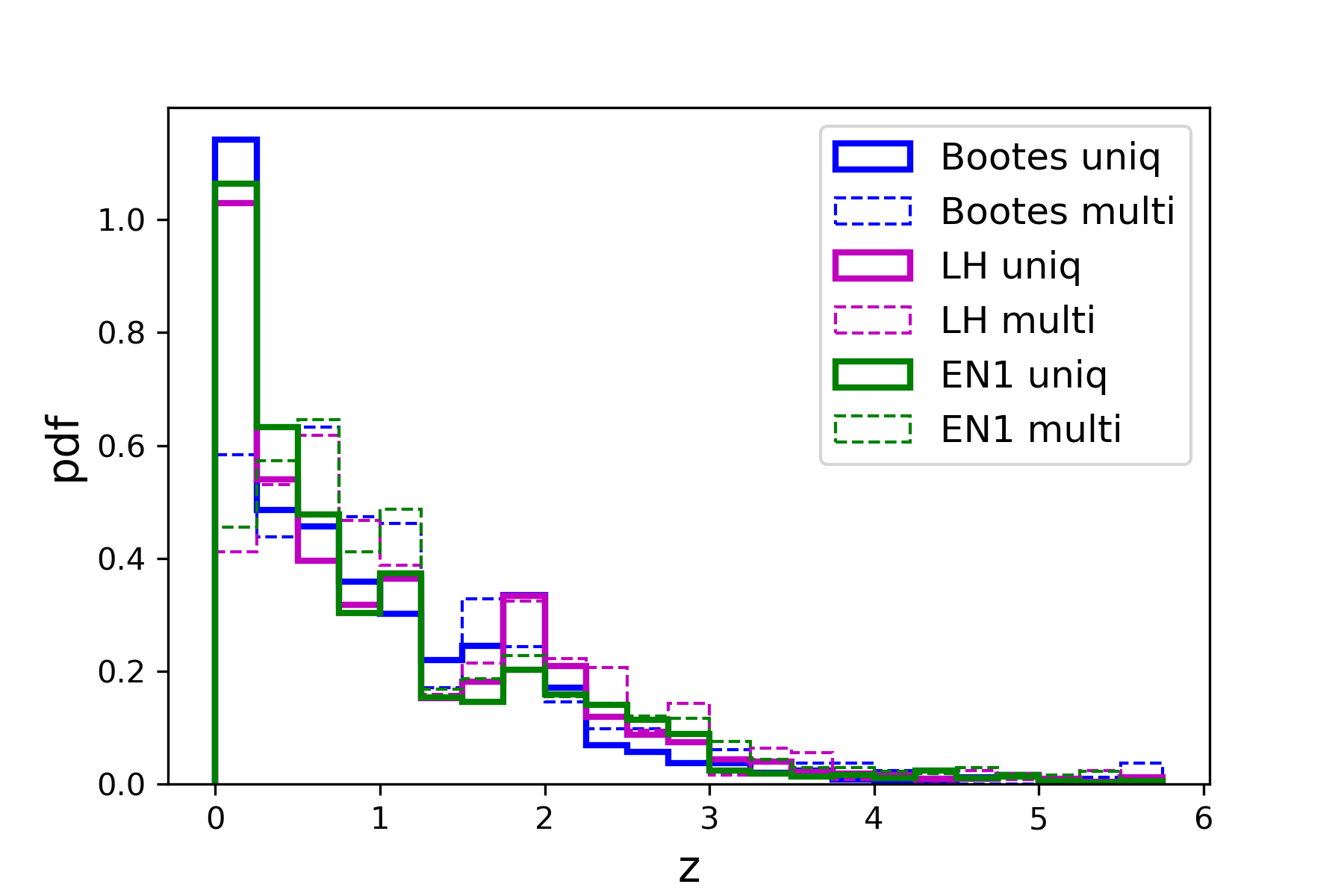}
\caption{Normalised redshift distributions (including both spectroscopic and photometric redshifts) of our {\it Herschel}-LOFAR matched samples in the LoTSS Deep Fields (colour-coded by field). The unique and multiple matches are shown by the solid and dashed lines, respectively.}
\label{Nz}
\end{figure}

\begin{table*}
\caption{Number statistics of the matched {\it Herschel}-LOFAR samples.  $S^{\rm limit}_{250\mu m}$ is the adopted 250 $\mu m$ flux density limit. $N^{\rm uni}_{250\mu m}$ is the number of {\it Herschel} sources with a unique LOFAR counterpart, and the number within the brackets is the number of these LOFAR counterparts with redshift estimates (including both spectroscopic and photometric redshifts). $N^{\rm mul}_{250\mu m}$ is the number of {\it Herschel} sources that are matched to multiple LOFAR sources. $N^{\rm mul}_{\rm 150MHz}$ is the number of LOFAR sources in these multiple matches, and the number inside the brackets indicates the number of these LOFAR sources with redshift estimate. $F^{\rm mul}$ is the fraction of {\it Herschel} sources that are matched to multiple LOFAR counterparts. $N^{\rm HLIRG}$ is the total number of HLIRGs, and $D^{\rm HLIRG}$ is the observed surface density of the HLIRGs.}\label{table:statistics}
\begin{tabular}{lllllllll}
\hline
\hline
Field & area (deg$^2$) & $S^{\rm limit}_{250\mu m}$ (mJy) & $N^{\rm uni}_{250\mu m}$ &   $N^{\rm mul}_{250\mu m}$  & $N^{\rm mul}_{\rm 150MHz}$ & $F^{\rm mul}$ & $N^{\rm HLIRG}$ & $D^{\rm HLIRG}$ (deg$^{-2}$)\\
\hline
Bo\"otes &  8.63 & 45 &  982 (974) &         178 &   330 (326) & 15\%& 44& 5.1\\
\hline
Lockman Hole (LH) &   10.28 & 40 & 1750 (1742)&        269 & 509 (507) & 13\%& 164& 16.0\\
\hline
Elais-N1 (ELAIS-N1) &  6.74 & 35  & 1484 (1470) &   654 &   1266 (1255) & 30\%&125&18.5\\ 
\hline
\hline
\end{tabular}
\end{table*}

\subsection{Matching statistics}

Table 1 lists  the number statistics of the matched {\it Herschel}-LOFAR samples in the Deep Fields. In Fig.~\ref{int_counts}, we plot the integral number counts of the DMU22 blind catalogues at 250, 350, and 500 $\mu$m in ELAIS-N1 and compare them with the counts of the unique and multiple matches. We note that we applied our flux limit at 250 $\mu m$, which is why there are no data points below 35 mJy for the raw counts from DMU22 and for the counts of the unique matches. Most of the bright {\it Herschel} sources are matched to LOFAR sources. Furthermore, most of the {\it Herschel}-LOFAR matches are unique, with a single LOFAR counterpart: 70\% in ELAIS-N1 (above the $S_{250\mu m}>35$ mJy limit), 87\% in LH ($>40$ mJy), and 85\% in Bo\"otes ($>45$ mJy). In Fig.~\ref{Nz}, we plot the normalised redshift distributions of the {\it Herschel}-LOFAR matches. The redshift distributions are fairly similar across the three fields. Above $z\sim0.25$, the unique and multiple redshift distributions are similar -- the difference is a large extra number of unique sources in each field at low-$z$. The median redshifts of the unique matches are 0.67, 0.77, and 0.68 in Bo\"otes, LH, and ELAIS-N1, respectively. In comparison, the median redshift of the multiple matches is 0.92, 0.97, and 0.94 in Bo\"otes, LH, and ELAIS-N1, respectively.

\subsubsection{The issue of multiplicity}

It is possible that the true level of multiplicity among the bright {\it Herschel} sources is higher than indicated by the numbers above (13 - 30\% depending on the $S_{250 \mu m}$ limit) as the contributing galaxies could be closer than the LOFAR beam FWHM $\sim6\arcsec$.  We checked 11 HLIRGs in the LH field that fall in the area within which high-resolution LOFAR imaging has been carried out using the international stations (Sweijen et al, 2020, in prep.). Two sources show evidence of AGN lobes, and the rest just show a single unresolved source down to the resolution of 0.3\arcsec, which seems to corroborate the low level of multiplicity found in our bright {\it Herschel}-selected samples. Additionally, as we discussed in Sect. 3.2, the current depth of the LoTSS Deep Fields allows us to detect multiples with flux density ratios up to 3 - 4 for the faintest 250 $\mu m$ sources in our samples. Therefore, galaxies that are more than 3 - 4 times fainter than the applied $S_{250 \mu m}$ limit could be missed. 

The multiplicity level of FIR- and sub-mm-selected sources is still an important issue to address. Follow-up high-resolution observations of bright SCUBA-2 850 $\mu m$ sources and {\it Herschel}-selected sources with interferometric facilities -- such as the Atacama Large Millimeter/submillimeter Array (ALMA; Wootten \& Thompson 2009), the Plateau de Bure Interferometer (PdBI; Guilloteau et al. 1992), and the Submillimeter Array (SMA; Ho, Moran \& Lo 2004) -- have revealed that some of these bright single-dish sources can indeed be resolved into multiple galaxies (as a result of chance projection or true physical associations). However, the multiplicity rate varies a lot, from 15-20\% to $\sim$70\% (e.g. Karim et al. 2013; Hodge et al. 2013; Simpson et al. 2015; Bussmann et al. 2015; Stach et al. 2018; Hatziminaoglou et al. 2018), depending on factors such as the beam-size of the single-dish telescope, sample selection, and the exact definition of multiplicity. 

Different definitions of a multiple exist in the literature and depend on the sensitivity limit applied to the secondary components. For example, Lambas et al. (2012) used flux density ratios to define multiples and considered multiples as pairs in which the brightest galaxy is less than three times brighter than the second-brightest galaxy. Much higher flux density ratios have also been used. For example, Bussmann et al. (2015) defined multiplicity using individual components that are up to a factor of $\sim15$ fainter than the total flux density of a given detection. Stach et al. (2018) considered secondary components up to a flux density ratio of $\sim9$ between the brightest and second-brightest galaxies. If the definition of multiples is restricted to the flux density ratio between the brightest and secondary components $< 2$ - 3 (which represent the single-dish detections most seriously affected by multiplicity), then there seems to be a consensus that the level of multiplicity is quite low. For example,  Micha{\l}owski et al. (2017) determined that only 15 - 20\% of the 850-$\mu m$ sources selected with SCUBA-2/JCMT (which has a similar beam size to {\it Herschel} at 250 $\mu m$) are significantly affected by multiplicity. Hill et al. (2018) observed the brightest sources in the SCUBA-2 Cosmology Legacy Survey (S2CLS) using the SMA and find that $<$ 15\% of the sources resolve into multiple galaxies. Fudamoto et al. (2017), using the IRAM NOrthern Extended Millimeter Array (NOEMA) and ALMA, observed 21 galaxies selected from the {\it Herschel} Astrophysical Large Area Survey (H-ATLAS; Eales et al. 2010), which are expected to be unlensed and extremely luminous (with median $L_{\rm IR}$ $\sim10^{13} L_{\odot}$) at $z>4$ (Ivison et al. 2016). They  find that none of their sources are affected by blending down to sub-arcsec resolution. Neri et al. (2020) targeted 13 bright galaxies detected in H-ATLAS with $S_{500\mu m}>80$ mJy with NOEMA and find that the majority of targets are individual sources down to 1.2\arcsec.

\subsubsection{The issue of lensing magnification}

Gravitationally lensed, high-redshift galaxies can be efficiently selected by searching for bright sources in wide area blank-field sub-mm surveys (Blain 1996;  Negrello et al. 2007, 2010; Wardlow et al. 2013). As we are focusing on bright {\it Herschel} sources to study luminous IR galaxies in this study, it is important to discuss the probable fraction of strongly lensed galaxies in our samples as gravitational lensing can significantly increase the flux of the affected sources. It has been demonstrated both theoretically and observationally that a simple flux cut at 500 $\mu m$ can result in high efficiency at selecting lensed high-redshift galaxies, after excluding contaminants such as radio-loud blazars (de Zotti et al. 2005; Gonz{\'a}lez-Nuevo et al. 2010) and local ($z<0.1$) late-type galaxies (Dunne et al. 2000; Serjeant \& Harrison 2005). Based on a statistical model of the gravitational lensing of submillimetre galaxies (SMGs) by a distribution of foreground masses, Wardlow et al. (2013) show that 32 - 74\% of the sources with 500 $\mu m$ flux $S_{500 \mu m}>100$ mJy are strongly lensed and the fraction of strongly lensed sources decreases to 14 - 40\% at $S_{500 \mu m}>80$ mJy. In our {\it Herschel}-LOFAR cross-matched samples, fewer than 1\% of the sources have $S_{500 \mu m}>80$ mJy. Therefore, we conclude that our samples of bright {\it Herschel} sources  are probably not significantly affected by lensing magnification.

\section{The bright end of the IR luminosity functions}

\begin{figure*}
\includegraphics[height=4.3in,width=7.6in]{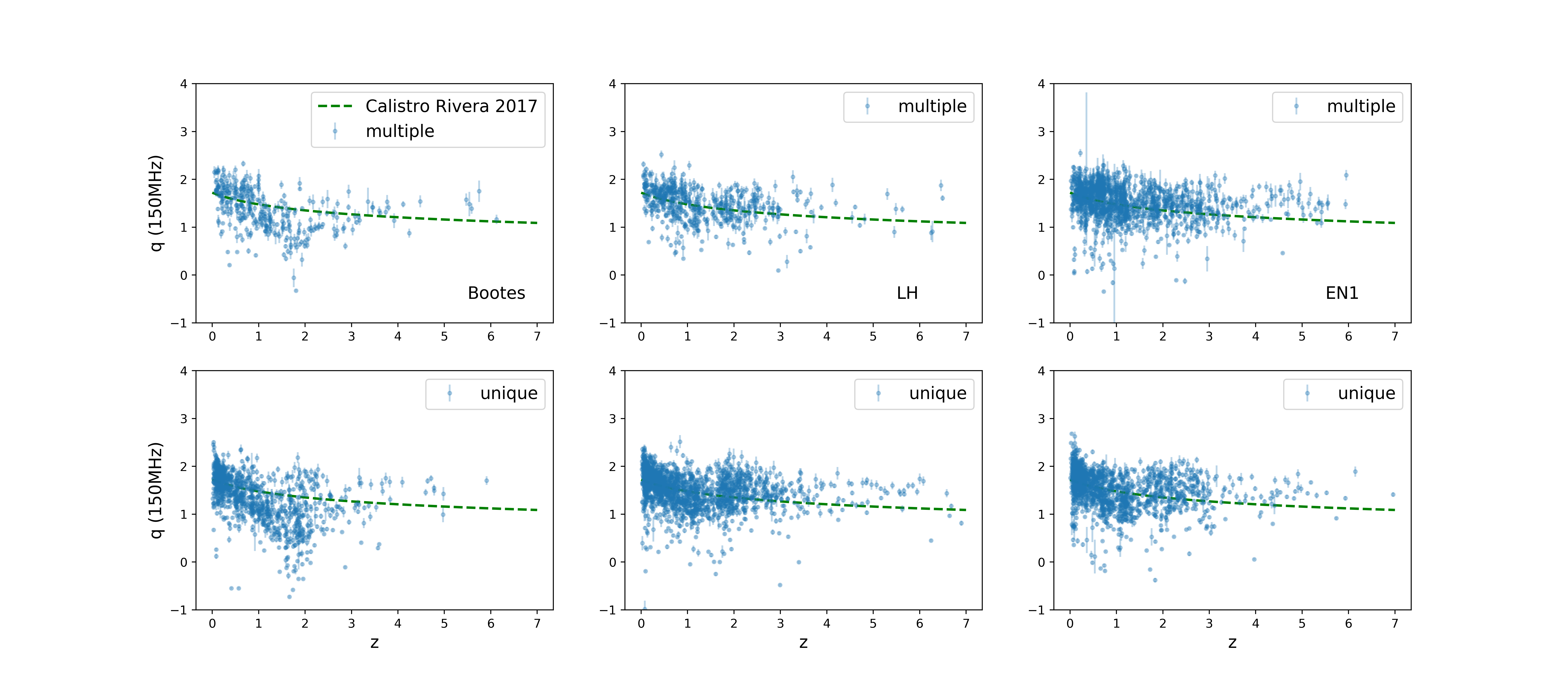}
\caption{Value of $q$  as a function of redshift in Bo\"otes, Lockman, and ELAIS-N1 from the multiple and  unique {\it Herschel}-LOFAR matches. The Calistro Rivera et al. (2017)  relation is shown as the green dashed line.}
\label{qirc}
\end{figure*}

\begin{figure*}
\includegraphics[height=9in,width=7.in]{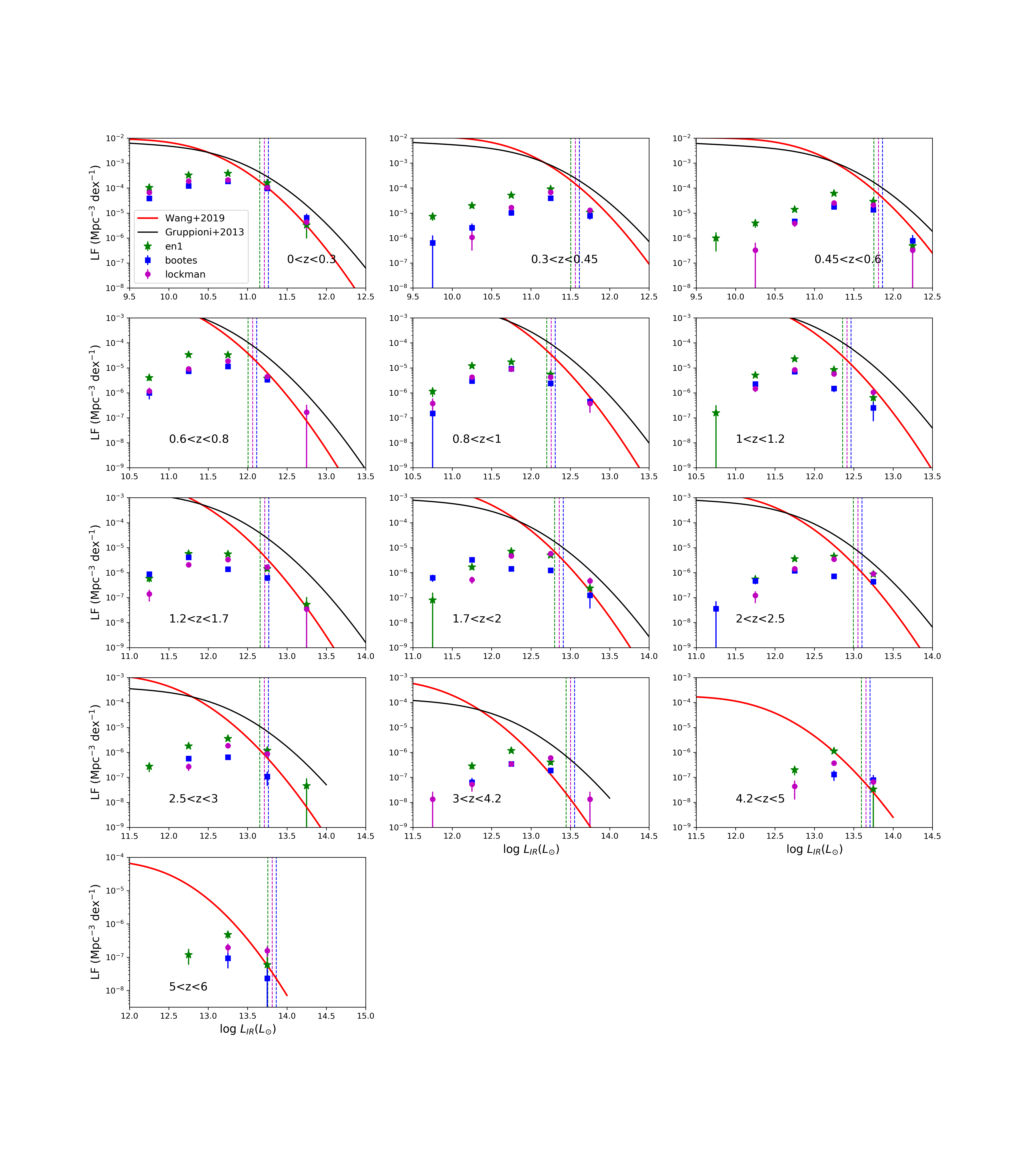}
\caption{Volume density of our {\it Herschel}-LOFAR matches (by combining unique and multiple matches)  as a function of IR luminosity in the three fields from $z\sim0$ to 6 (colour-coded by field), compared to LFs from Wang et al. (2019a) and Gruppioni et al. (2013). The vertical dashed line indicates the corresponding luminosity completeness limit in each field.}
\label{en1_LF}
\end{figure*}

\subsection{Estimate IR luminosity using the CIGALE SED fitting}

We used the SED modelling and fitting tool Code Investigating GALaxy Emission (CIGALE; Burgarella et al. 2005; Noll et al. 2009; Serra et al. 2011; Boquien et al. 2019) to fit the multi-wavelength photometric data, from ultraviolet (UV) to IR, of the galaxies cross-matched between {\it Herschel} and LOFAR. The CIGALE tool uses an energy balance approach between the attenuated UV and optical emission and the re-emitted IR and sub-mm emission. The choices for the SED model components follow Pearson et al. (2018). Briefly, we used a delayed exponentially declining star-formation history (SFH) with an exponentially declining burst, Bruzual \& Charlot (2003) stellar emission, Chabrier (2003) initial mass function, Charlot \& Fall (2000) dust attenuation, the updated Draine et al. (2014) version of the Draine \& Li (2007) IR dust emission, and Fritz et al. (2006) AGN models. Using CIGALE, we could therefore estimate the $L_{\rm IR}$ of the bright {\it Herschel} galaxies in our samples.

In order to assess the level of agreement in the $L_{\rm IR}$ estimates among different SED fitting codes, we also compared our CIGALE-derived $L_{\rm IR}$ with estimates derived using MAGPHYS (da Cunha et al. 2008), AGNfitter (Calistro Rivera et al. 2016), and a new version of CIGALE with a new AGN model, SKIRTOR (Stalevski et al. 2012, 2016).  All of these codes were run on the $\sim30,000$ LOFAR source host galaxies in ELAIS-N1 as part of the process to classify the sources (star-forming galaxies versus different classes of AGNs) and to derive source characteristics (Best et al. 2020). We find that the mean difference in $L_{\rm IR}$  is 0.167 dex between CIGALE and MAGPHYS, 0.281 dex between CIGALE and AGNfitter, and 0.195 dex between the CIGALE run with the SKIRTOR model and the CIGALE run with the Fritz model. Therefore, we can conclude that there is generally a good level of agreement (with small systematic offsets) among the different SED fitting codes.

\subsection{Results}

Star-forming galaxies are found to obey the FIRC as UV photons from young stars heat dust grains that then radiate in the IR, and the same short-lived massive stars explode as supernovae, which accelerate cosmic rays, thereby contributing to non-thermal synchrotron emission in the radio. Therefore, we also checked the $q$ parameter of our {\it Herschel}-LOFAR matched samples. Because the FIRC has a slope of near unity, it is common to use the $q$ parameter, which can be defined as (Bell 2003)
\begin{equation}
q = \log \left(   \frac{L_{\rm IR}/(3.75\times10^{12}{\rm Hz})}{L_{150 {\rm MHz}}}  \right),
\end{equation}
where $L_{\rm IR}$ is the rest-frame IR luminosity in units of erg s$^{-1}$ and $L_{150 {\rm MHz}}$ is the 150-MHz luminosity in erg s$^{-1}$ Hz$^{-1}$.  Figure~\ref{qirc} shows the $q$ value as a function of redshift for the unique and multiple matches. There is a good level of agreement between the multiple and unique matches as well as across the three fields. The dashed line represents the best-fit relation derived by Calistro Rivera et al. (2017) using deep LOFAR observations in the Bo\"otes field (Williams et al. 2016). The Calistro Rivera et al. (2017) relation has  a functional form of $q \propto (1+z)^{\gamma}$, with $\gamma=-0.15\pm-0.032$, and it seems to provide a good fit to our data. It is clear that the FIRC holds for the majority of the sources in our samples, which further supports the reliability of our {\it Herschel}-LOFAR matches and implies that our photo-$z$ estimates are reasonable. A small number of sources (expected to be radio-loud AGNs) show excess radio emission. At face value, both our results and Calistro Rivera et al. (2017) suggest that there is a mild redshift evolution. However, as we pointed out previously (Wang et al. 2019b), the apparent evolution could be a consequence of the non-linearity of the FIRC  (i.e. non-unity slope) and the fact that galaxies are more luminous at higher redshifts. McCheyne et al. (in prep.) will present a detailed study of the FIRC using data from the LoTSS Deep Fields.

The LF simply measures the volume density of galaxies as a function of their intrinsic luminosity and can provide strong constraints on models of galaxy formation and evolution (e.g. Granato et al. 2004; Baugh et al. 2005; Fontanot et al. 2007; Hayward et al. 2013; Cowley et al. 2015; Lacey et al. 2016). Figure~\ref{en1_LF} show the IR LFs from the local Universe all the way to $z\sim6$ in the three Deep Fields and how they compare with earlier {\it Herschel}-based results from Gruppioni et al. (2013) and Wang et al. (2019a). Gruppioni et al. (2013) used the data  from the {\it Herschel} PEP Survey at 70, 100, and 160 $\mu m$ with the HerMES data at 250, 350, and 500 $\mu m$ to derive the IR LFs up to $z\sim4$. Wang et al. (2019a) used a state-of-the-art de-blended {\it Herschel} catalogue in the COSMOS field to derive the IR LF out to $z\sim6$. The solid black and red  lines in Fig.~\ref{en1_LF} show the best-fit LFs from Gruppioni et al. (2013) and Wang et al. (2019a), respectively, which are characterised by a modified Schechter function (e.g. Saunders et al. 1990; Wang \& Rowan-Robinson 2010; Marchetti et al. 2016; Wang et al. 2016) to fit the total IR LF,
\begin{equation}
\phi(L) = \phi_* \left(\frac{L}{L_*}\right)^{1-\alpha} \exp\left[ - \frac{1}{2\sigma^2} \log^2_{10}\left(1+\frac{L}{L_*}\right)\right].
\end{equation}
The vertical dashed line indicates the corresponding luminosity completeness limit in each field, which will be discussed in Sect. 5. At the bright end of the LF, our results show excellent agreement with the LFs presented by Wang et al. (2019a). In comparison, the LFs presented in Gruppioni et al. (2013) are systematically higher compared to both the findings of this study and those of Wang et al. (2019a).

\section{The abundance of hyperluminous infrared galaxies (HLIRGs)}

\begin{figure}
\includegraphics[height=2.5in,width=3.4in]{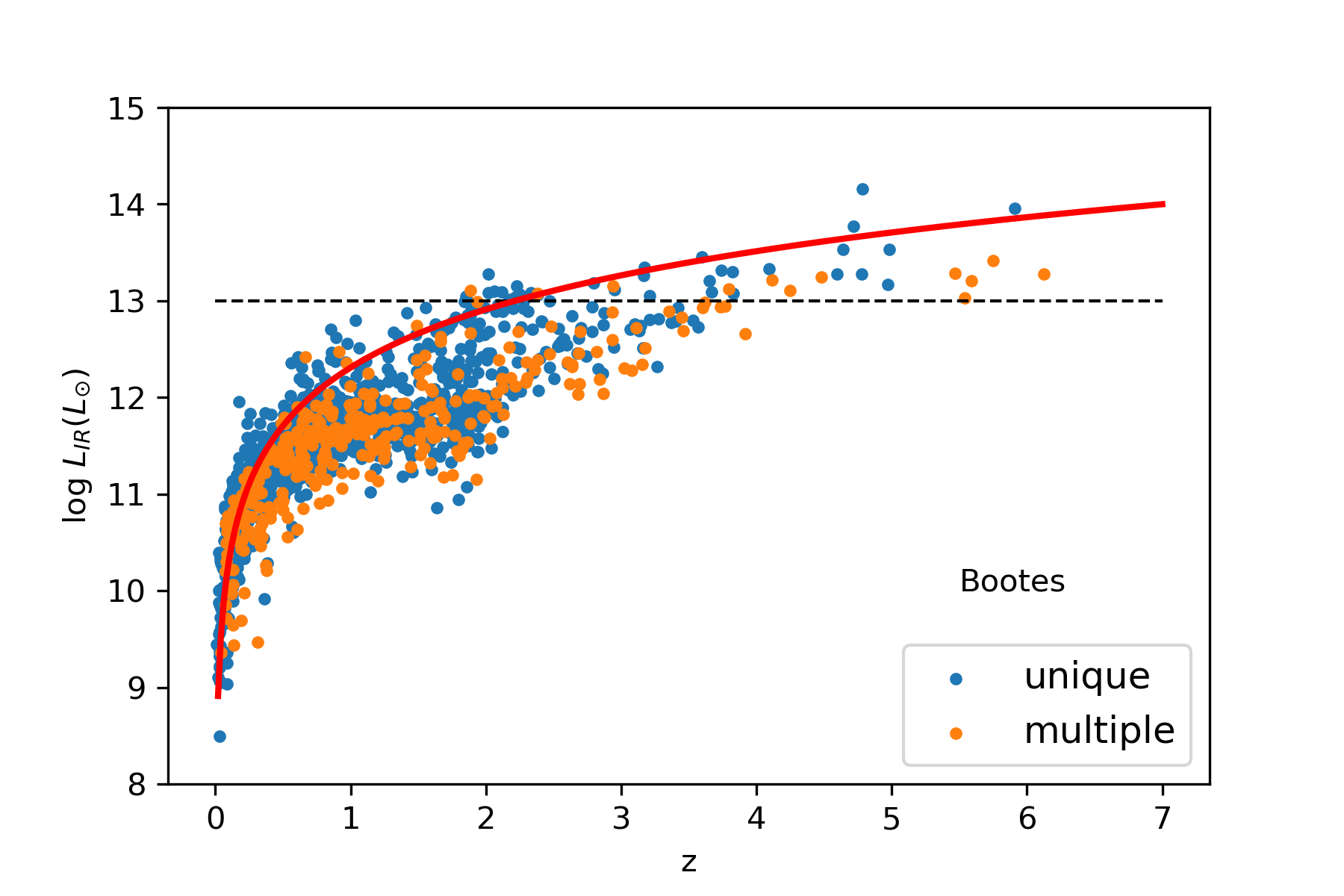}
\includegraphics[height=2.5in,width=3.4in]{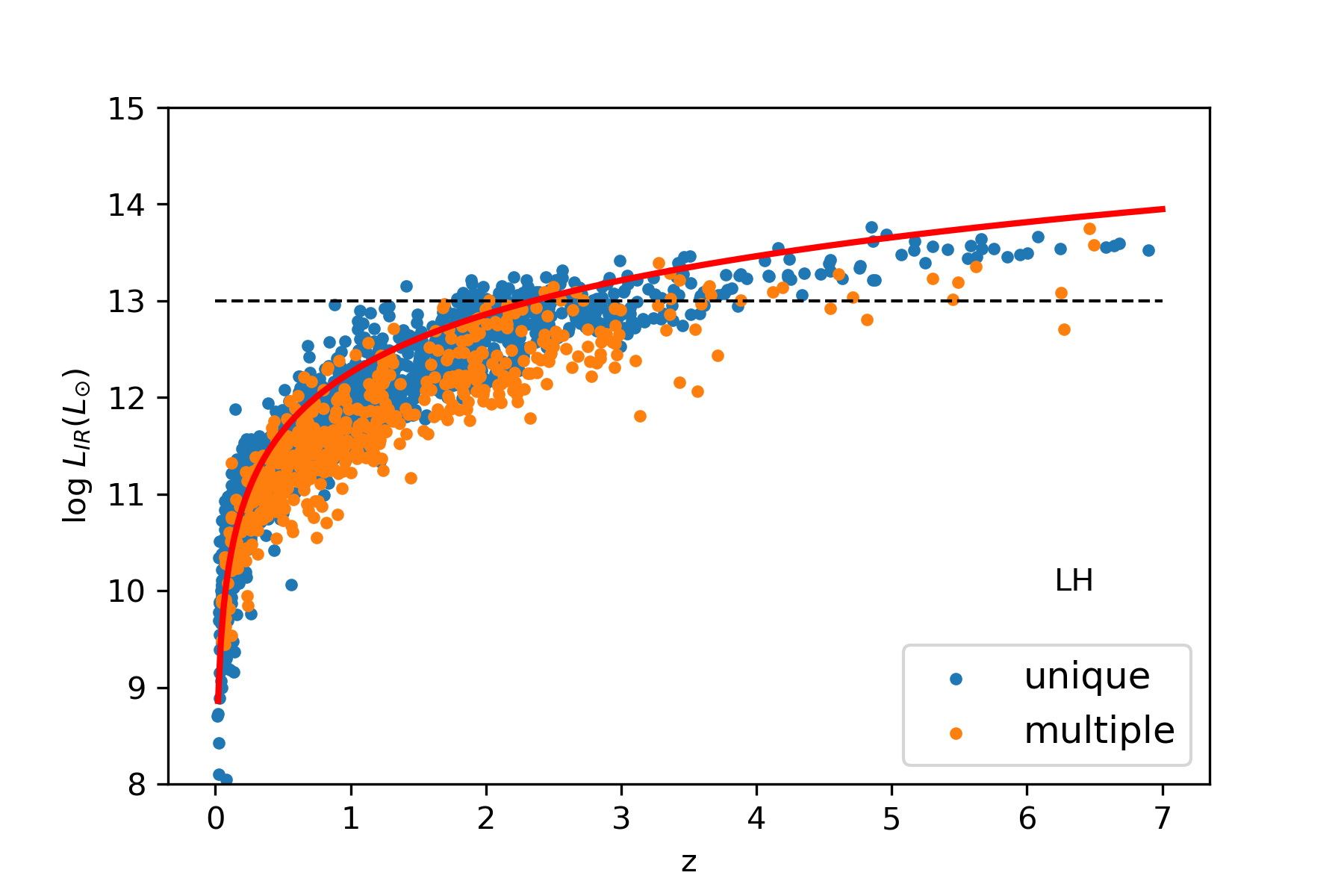}
\includegraphics[height=2.5in,width=3.4in]{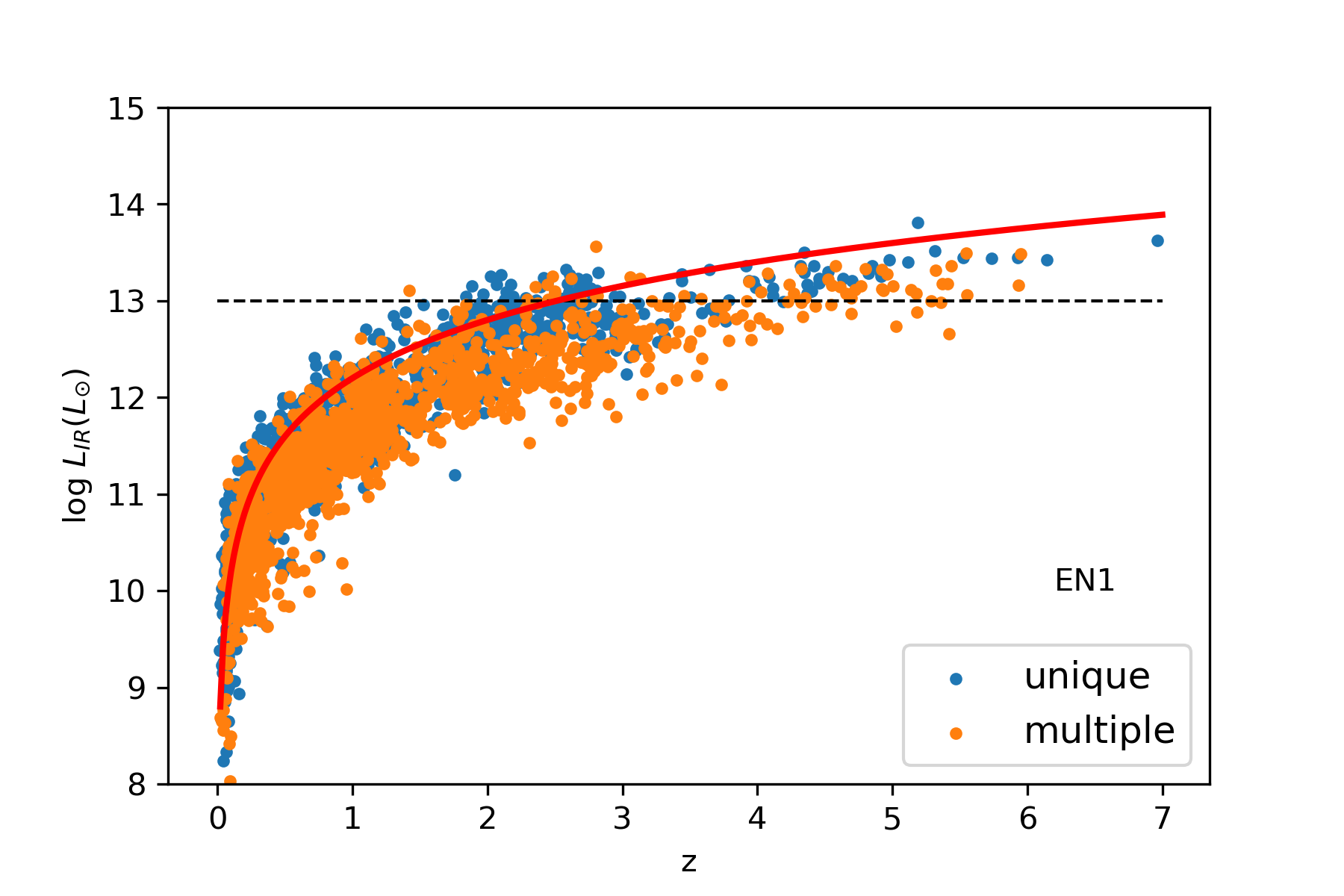}
\caption{IR luminosity $L_{\rm IR}$ vs. redshift (including both spectroscopic and photometric redshifts) of our {\it Herschel}-LOFAR cross-matched samples in the three Deep Fields, colour-coded by unique or multiple matches.  The horizontal dashed line indicates the luminosity of the most luminous IR galaxy found to date. The solid red line represents the completeness limit on $L_{\rm IR}$ as a function of redshift in each field.}
\label{LIR_z}
\end{figure}

\begin{figure*}
\includegraphics[height=2.7in,width=3.6in]{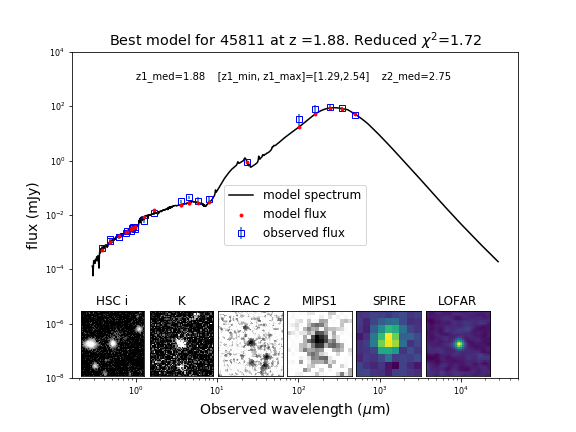}
\includegraphics[height=2.7in,width=3.6in]{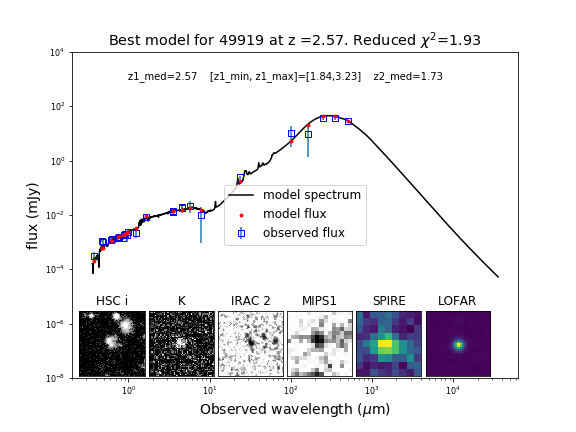}
\includegraphics[height=2.7in,width=3.6in]{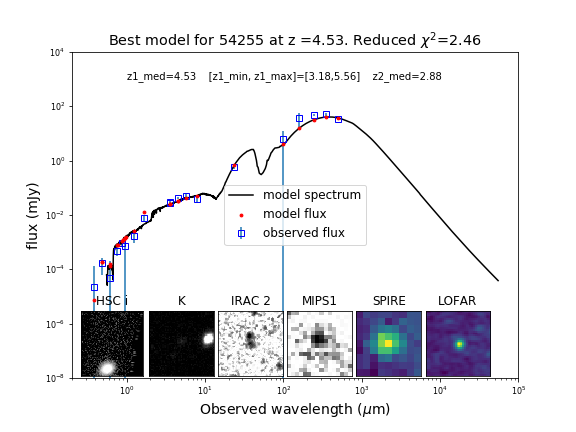}
\includegraphics[height=2.7in,width=3.6in]{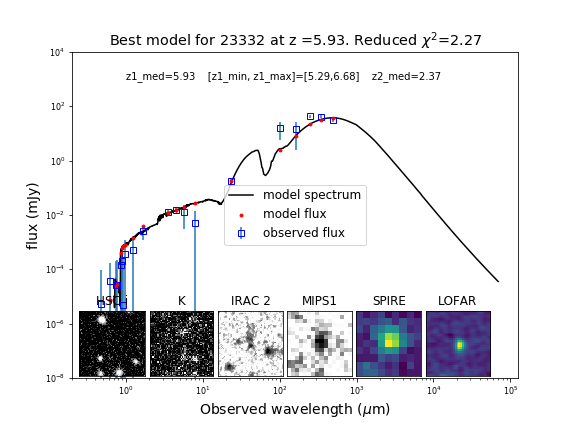}
\caption{Example SED fits of HLIRGs in the unique category in ELAIS-N1. In each panel, z1\_med corresponds to the primary redshift peak in the probability density function, [z1\_min, z1\_max] corresponds to the range between the lower and upper bounds of the primary 80\% HPD CI peak, and  z2\_med corresponds to the secondary redshift peak (if present). Image cutouts from the optical to radio (from left to right, HSC $i$-band 10.8\arcsec $\times$ 10.8\arcsec, DXS K-band 10.2\arcsec $\times$ 10.2\arcsec, IRAC 4.5 $\mu m$ 36.0\arcsec $\times$ 36.0\arcsec, MIPS 24 $\mu m$ 36.0\arcsec $\times$ 36.0\arcsec, SPIRE 250 $\mu m$ 54.0\arcsec $\times$ 54.0\arcsec,  and LOFAR 150 MHz 54.0\arcsec $\times$ 54.0\arcsec) are shown in a row of insets at the bottom.}
\label{SED_IR1}
\end{figure*}

\begin{figure*}
\includegraphics[height=2.7in,width=3.6in]{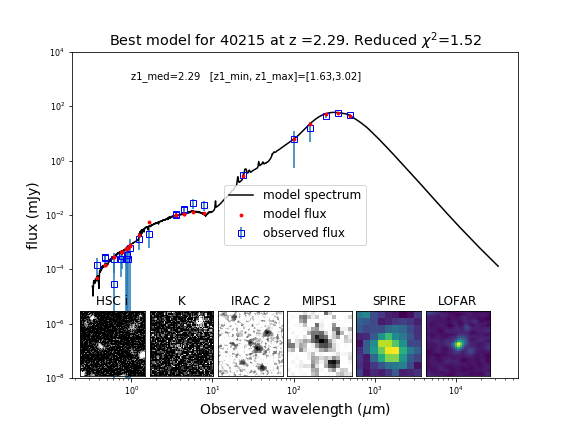}
\includegraphics[height=2.7in,width=3.6in]{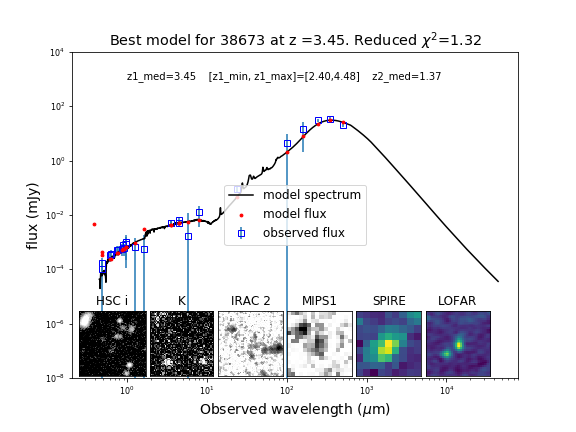}
\includegraphics[height=2.7in,width=3.6in]{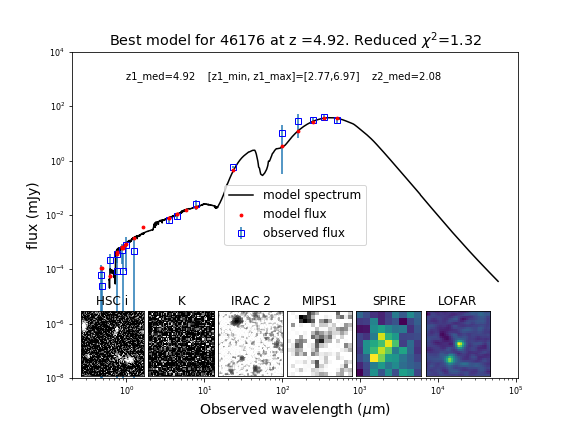}
\includegraphics[height=2.7in,width=3.6in]{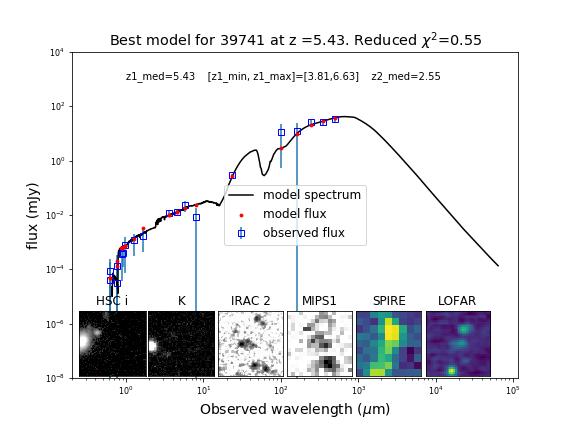}
\caption{Example SED fits of HLIRGs in the multiple category in ELAIS-N1. In each panel, z1\_med corresponds to the primary redshift peak in the probability density function, [z1\_min, z1\_max] corresponds to the range between the lower and upper bounds of the primary 80\% HPD CI peak, and  z2\_med corresponds to the secondary redshift peak (if present). Image cutouts from the optical to radio (from left to right, HSC $i$-band 10.8\arcsec $\times$ 10.8\arcsec, DXS K-band 10.2\arcsec $\times$ 10.2\arcsec, IRAC 4.5 $\mu m$ 36.0\arcsec $\times$ 36.0\arcsec, MIPS 24 $\mu m$ 36.0\arcsec $\times$ 36.0\arcsec, SPIRE 250 $\mu m$ 54.0\arcsec $\times$ 54.0\arcsec, and LOFAR 150 MHz 54.0\arcsec $\times$ 54.0\arcsec) are shown in a row of insets at the bottom.}
\label{SED_IR2}
\end{figure*}

\begin{figure*}
\includegraphics[height=2.5in,width=7.8in]{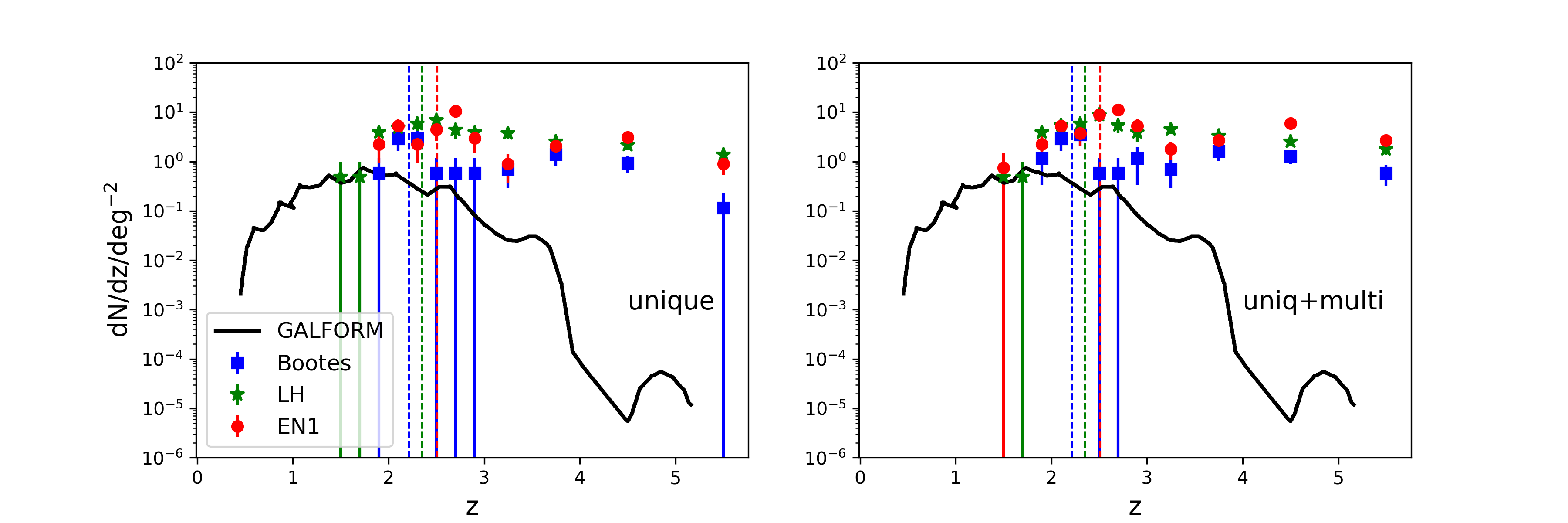}
\caption{Observed surface density of HLIRGs (with Poisson errors) as a function of redshift in the three fields (left: unique matches; right: unique and multiple matches), compared with the predicted surface density from GALFORM. The vertical dashed lines represent the redshift boundaries below which the HLIRG sample is complete for each field. }
\label{dndz}
\end{figure*}

In Fig.~\ref{LIR_z}, we plot $L_{\rm IR}$ versus redshift for the {\it Herschel}-LOFAR matches in the LoTSS Deep Fields. The most luminous galaxy in our sample has $L_{\rm IR} = 1.4\times10^{14} L_{\odot}$ and is the only galaxy with $L_{\rm IR}>10^{14}L_{\odot}$. 

To work out the $L_{\rm IR}$ limit as a function of redshift imposed by the 250 $\mu m$ flux limit (35 mJy in ELAIS-N1; 40 mJy in LH; 45 mJy in Bo\"otes), and taking into account the variations in the IR SED shapes, we first checked the distribution of sources with $S_{250\mu m}$ close to our selection limit in the $L_{\rm IR}$ versus $z$ parameter space. For example, in Bo\"otes, we examined the distribution of sources with $44<S_{250\mu m}<46$ mJy in $L_{\rm IR}$ versus $z$. The upper boundary of this distribution was then designated as the completeness limit on $L_{\rm IR}$ as a function of redshift, as this is the largest $L_{\rm IR}$ for a source at our selection limit at a given redshift. The solid red line represents the derived completeness limit on $L_{\rm IR}$ as a function of redshift in each field. In our deepest field, ELAIS-N1, we are complete in the selection of ULIRGs out to $z\sim0.8$ and HLIRGs out to $z\sim2.5$.

In all three fields, we can see examples of HLIRGs with $L_{\rm IR} > 10^{13} L_{\odot}$ at $z>1$ in both the unique and multiple categories. The LH field has the largest HLIRG population, which is due to the larger area of the field compared to ELAIS-N1 and Bo\"otes. The total numbers of  HLIRGs are:  76 (unique) + 49 (multiple) = 125  in ELAIS-N1; 32 (unique) + 12 (multiple) = 44 in Bo\"otes; and 138 (unique) + 26 (multiple) =  164 in LH. Therefore, the observed surface densities of the HLIRGs are 18.5, 16.0, and 5.1 per deg$^2$ in ELAIS-N1, LH, and Bo\"otes, respectively. In order to test the impact of the varying flux limit across the three fields, we also checked the number of HLIRGs by raising the flux limit to 45 mJy in ELAIS-N1 and LH. Thresholding at the same flux limit of 45 mJy as the limit applied in Bo\"otes, the total number of HLIRGs is reduced to 43 (unique) + 27 (multiple) = 70  in ELAIS-N1 and 105 (unique) + 18 (multiple) =  123 in LH, leading to observed surface densities of 10.4 and 12.0 per deg$^2$ in ELAIS-N1 and LH, respectively. Therefore, the lower observed surface density of HLIRGs in Bo\"otes is caused by the higher flux limit at 250 $\mu m$ as well as cosmic variance. The number statistics of the HLIRG samples is summarised in Table 1. 

Table 2 shows the number statistics of AGNs identified in the HLIRG samples in the three fields. Infrared AGNs are identified using the {\it Spitzer} IR  colour and selection criteria from Donley et al. (2012). Optical AGNs are identified via cross-matching with the Million Quasar Catalog based on the  Sloan Digital Sky Survey (SDSS; Alam et al. 2015) and other literature catalogues (Flesch 2015). In the Bo\"otes field, $X$-ray AGNs are identified via cross-matching with the X-Bo\"otes {\it Chandra} survey of the NOAO Deep Wide-Field Survey (NDWFS; Kenter et al. 2005). In ELAIS-N1 and LH, X-ray AGNs are identified via cross-matching with $X$-ray sources detected in the Second {\it ROSAT} All-Sky Survey (2RXS; Boller et al. 2016) and the {\it XMM-Newton} Slew Survey (XMMSL2). We note that the $X$-ray data are deeper in the Bo\"otes field than in ELAIS-N1 and LH. Further details on AGN identification can be found in Duncan et al. (2020). In Figs.~\ref{SED_IR1} and ~\ref{SED_IR2}, we show example SED fits to HLIRGs (in the unique and multiple categories) in ELAIS-N1 and image cutouts at the Hyper Suprime-Cam (HSC) $i$-band, the UKIDSS Deep eXtra-Galactic (DXS) K-band, the Infrared Array Camera (IRAC) 4.5 $\mu m$, the Multiband Imaging Photometer (MIPS) 24 $\mu m$, SPIRE 250 $\mu m$, and LOFAR 150 MHz.  The HLIRGs tend to be extremely faint in the optical and NIR and only detected at high S/N in the IRAC bands and at longer wavelengths. Many HLIRGs are extreme starbursts with SFRs of more than 1000 $M_{\odot}$/yr. A follow-up study by Gao et al. (in prep.) will examine the physical properties of these extraordinary galaxies, the power source, their environment, and their contribution to the cosmic SFH  in detail.

\begin{table*}
\caption{Number statistics of AGNs identified in the HLIRG samples. N (Opt AGN) represents the number of optical AGNs identified. N(IR AGN) represents the number of IR AGNs. N ($X$-ray AGN) represents the number of $X$-ray AGNs. N (AGN) is the total number of AGNs identified in the optical, IR, or $X$-ray. Frac (AGN) is the fraction of AGNs in the HLIRG samples.}\label{table:HLIRGstatistics}
\begin{tabular}{llllll}
\hline
\hline
Field & N (IR AGN) & N (Opt AGN) & N ($X$-ray AGN) & N (AGN) & Frac (AGN)\\
\hline
Bo\"otes & 10 & 2 & 1 & 11 & 25.0\%\\
\hline
Lockman Hole (LH) & 42 & 2& 0& 43 & 26.2\%\\
\hline
Elais-N1 (ELAIS-N1) & 20 & 1& 0& 20& 16.0\%\\ 
\hline
\hline
\end{tabular}
\end{table*}

Although the photometric redshifts have been extensively calibrated and tested, and the fact that the FIRC holds for our sample corroborate that the photo-$z$ estimates are reasonable, we wished to further test the influence of photo-$z$ quality (which generally decreases towards higher redshift) on the abundance of HLIRGs using three different approaches. In the first approach, we applied a cut on the width of the primary redshift peak in the photo-$z$ probability distribution to select good-quality photo-$z$. Specifically, we used $\delta z = (z1\_max - z1\_min)/2$ as a metric on the photo-$z$ quality. Here, $z1\_min$ ($z1\_max$) is the lower (upper) bound of the primary 80\% highest probability density (HPD) credible interval (CI) peak. We required $\delta z/(1 + z)<0.15$ to select good-quality photo-$z$. We find that by restricting our HLIRGs sample to sources with 'good' photo-$z$, the observed surface density is reduced by a factor of $\sim2$. In the second approach, we re-ran our CIGALE SED fitting routine to derive $L_{\rm IR}$, but this time using $z2$\_median (the median of the secondary redshift peak above 80\% HPD CI), if present, rather than $z1$\_median (the median of the primary redshift peak) for sources without spectroscopic redshifts. In ELAIS-N1, we find the total number of HLIRGs changes from 125 to 169. Therefore, we observe an increase of $\sim35\%$ in the density of HLIRGs if the secondary redshift peak are used. Finally, we also tested whether the LOFAR photo-$z$, which are derived using optical, NIR, and IRAC data, are consistent with the IR SEDs. The IR SED can provide an independent estimate of the photo-$z$ (albeit with much lower precision) since the MIR to FIR emission has a generally well-defined peak at 50 - 200 $\mu m$. We find a mean difference of 0.085 (implying a low level of systematic offset), which indicates that the LOFAR photo-$z$ are, in general, consistent with the IR SEDs.

In Fig.~\ref{dndz}, we compare the observed surface density of HLIRGs as a function of redshift with predictions from the Durham semi-analytic models of galaxy formation, GALFORM. In GALFORM, galaxies populate dark matter halo merger trees according to simplified prescriptions for the baryonic physics involved. Here, we used the version of GALFORM presented by Lacey et al. (2016) with a minor re-calibration due to the model being implemented in an updated Planck cosmology (Baugh et al. 2019). This model is calibrated to reproduce a large set of observational data, including sub-mm galaxy number counts.  The predicted density for HLIRGs from GALFORM (derived from the mock lightcone catalogue) peaks at $z\sim2$, close to the peak in the cosmic SFH, and it declines rapidly towards both lower and higher redshifts. In comparison, the observed density for HLIRGs shows much less variation across the entire redshift range. There is a good level of agreement between ELAIS-N1 and LH, while Bo\"otes shows lower densities due to a higher flux cut at 250 $\mu m$ (and/or cosmic variance effects). The vertical dashed lines represent the redshift boundary below which the HLIRG sample is complete for each of the three Deep Fields. We note that we have not applied any correction factors to correct for incompleteness in our sample, so, strictly speaking, the observed densities at redshifts beyond the dashed lines represent lower limits. The predicted values fall below the observations, and the discrepancy between the model predictions and observations becomes increasingly larger towards higher redshifts. For HLIRGs, the predicted values are lower than the observed values by around one, two, and three orders of magnitude at $z\sim2$, $\sim3,$ and $\sim4$, respectively. Strong lensing and close galaxy pairs are not expected to play a significant role in the large discrepancy between the observed values and model predictions. As discussed in Sect. 3.3.2, based on the model presented in Wardlow et al. (2013), we expect the fraction of strongly lensed sources in our HLIRG samples to be much lower than 14 - 40\% as fewer than 1\% of our sources have 500 $\mu m$ flux densities $> 80$ mJy. It is possible that some sources in our HLIRG samples are close-pairs that appear as a single source in the {\it Herschel} imaging data as well as in the higher-resolution LOFAR imaging data. However, close-pairs cannot account for the orders of magnitude differences between the observations and GALFORM predictions. In a merger system involving two galaxies, the brightest member by definition will have $>50$\% of the luminosity of the whole system. The numbers of HLIRGs in our samples with $L_{\rm IR} > 2 \times 10^{13} L_{\odot}$ are 27, 43, and 11 in ELAIS-N1, LH, and Bo\"otes, respectively, leading to observed surface densities of 4.0, 4.2, and 1.3 /deg$^2$ in ELAIS-N1, LH, and Bo\"otes, respectively. Finally, the impact of the photo-$z$ quality on the abundance of HLIRGs is relatively small (a factor of $\lesssim2$ based on the three different tests described in this section) and insufficient to explain the discrepancy between the observations and model predictions.

\section{Conclusions}

In this paper, we have exploited the combined power of LOFAR deep surveys and {\it Herschel} deep extragalactic surveys in the best-studied northern extragalactic deep fields to provide the most accurate estimate yet on the bright end of the IR LFs out to $z\sim6$ and the abundance of HLIRGs with IR luminosities. The {\it Herschel} wavebands allow us to select dusty star-forming galaxies with minimal bias caused by the intrinsic variations of the IR SED shapes. On the other hand, the high sensitivity and angular resolution of the LOFAR 150-MHz continuum imaging data offer us the unique opportunity to robustly identify the multi-wavelength counterparts of the $\it Herschel$-selected sources as well as de-blend sources by virtue of the FIRC. This powerful combination of {\it Herschel} and LOFAR results in a large sample of 250 $\mu m$-selected sources with unprecedented number statistics and dynamic range in IR luminosity and volume.

We cross-matched LOFAR source catalogues and {\it Herschel}-SPIRE blind source catalogues in three extragalactic fields: Bo\"otes, LH, and Elais-N1. The resulting {\it Herschel}-LOFAR cross-matched samples are highly reliable (with a $\lesssim5$\% false identification rate) and complete ($\sim92$\% completeness) above 250 $\mu m$ flux limits of $S_{250\mu m}=$ 45, 40, and 35 mJy in Bo\"otes, LH, and Elais-N1, respectively. We divided the {\it Herschel}-LOFAR matches into unique and multiple matches. Most of the {\it Herschel}-LOFAR  matches are found in the unique category, that is, a single LOFAR counterpart within the {\it Herschel}-SPIRE beam, with only 13 - 30\% found in the multiple category, depending on the adopted 250 $\mu m$ flux limit. For the multiple matches, there is an excellent degree of correspondence between the radio emission on the LOFAR maps and FIR emission on the SPIRE maps. We ran the probabilistic Bayesian source extraction tool XID+ to de-blend the {\it Herschel}-SPIRE fluxes for the multiple matches, using the precise LOFAR source positions and 150-MHz flux densities as priors. Using the robust multi-wavelength counterpart identification of the LOFAR radio sources and photometric redshift estimates, we then performed CIGALE SED fittings to derive IR luminosity.

Our main conclusions are:
\begin{itemize}
  \item The resulting  {\it Herschel} 250 $\mu m$-selected samples in this paper are highly reliable and complete. The flux densities are not expected to be significantly affected by any residual multiplicity from sources not included in our analysis (at most 20 - 25\% for the faintest sources in our samples and to a lesser degree for the brighter sources).  
  \item We find a good agreement in IR LFs with the previous study by Wang et al. (2019a) over the luminosity ranges where our samples are complete. 
  \item Our samples give the strongest indication to date that a large population of HLIRGs with surface densities of $\sim5$ to $\sim18$ per deg$^2$ exists. Photometric redshift quality  can affect the surface density of HLIRGs, but we estimate that the effect is relatively small (a factor of $\lesssim2$ lower or higher). 
  \item Comparing our results to predictions from the GALFORM semi-analytic model, we find the model significantly under-predicts the surface density of HLIRGs. The discrepancy between the observed and predicted densities increases rapidly with increasing redshift. At $z\sim 2$, 3, and 4, the deficits are around one, two, and three orders of magnitude, respectively. 
\end{itemize}

Our sample of HLIRGs is an important dataset with which to understand the nature of these extraordinarily luminous objects and to test models of galaxy formation and evolution. There is a clear and urgent need for follow-up observations, such as: deep and high-resolution imaging to reveal the intrinsic level of multiplicity and signs of gravitational lensing (and constrain the level of lensing magnification); spectroscopic observations in the optical and/or millimetre to confirm the hyperluminous nature; and integral field unit (IFU) observations to probe the internal dynamics and triggering mechanism (e.g. mergers vs. secular processes).

\begin{appendix}

\section{De-blending {\it Herschel}-SPIRE fluxes using XID+}

As explained in Sect. 3.2, in this paper we used our SED prior enhanced XID+  to de-blend the bright {\it Herschel} sources that are matched to multiple LOFAR counterparts within the {\rm Herschel} beam (i.e. in the multiple category of the {\it Herschel}-LOFAR matches). Specifically, we estimated the expected IR fluxes and uncertainties in the SPIRE wavebands based on the observed 150-MHz fluxes and the LOFAR to {\it Herschel} flux ratios as a function of redshift. The flux priors, combined with the precise positional information from the LOFAR source, are then fed into XID+ to de-blend a circular region with a 60\arcsec\, radius centred around the position of the blind {\it Herschel} detection.

It is worth pointing out that MIPS, PACS, and SPIRE fluxes were measured for the optical counterparts to the LOFAR sources and provided in the first LoTSS Deep Fields data release. First, a cross-match between the LOFAR optical positions and the HELP catalogues was done  with a search radius of 0.5\arcsec. This distance was chosen as it corresponds to the positional uncertainty of the IRAC measurements from Spitzer, which were used to construct the prior list for HELP. If a HELP source was within this search radius, then any MIPS, PACS, or SPIRE fluxes were assigned to the LOFAR counterpart. If no HELP FIR source was within 0.5\arcsec\,, then XID+ was rerun on a circular region (30\arcsec\, radius for MIPS and PACS, and 60\arcsec\, radius for SPIRE) around the LOFAR counterpart. The prior list used was the same as that used in HELP, but the LOFAR counterpart was added onto the prior list as well. Finally, if the LOFAR radio source had no optical counterpart from the cross-matching, then XID+ was rerun around the source in the same manner but with the radio position added onto the prior list.

\begin{figure*}
\includegraphics[height=9.in,width=5.2in]{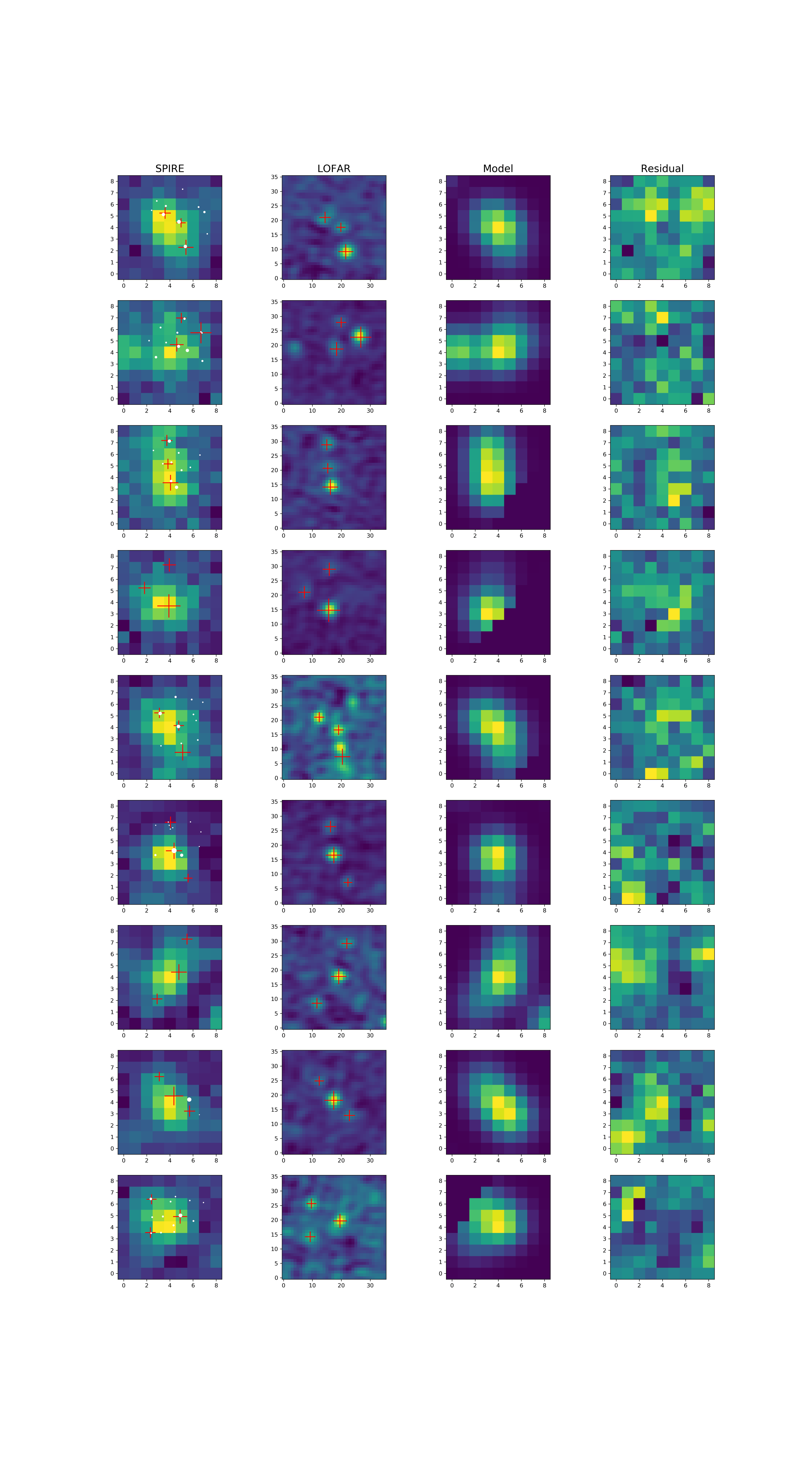}
\caption{From left to right, columns show postage stamps of the original {\it Herschel} 250 $\mu$m image, the LOFAR 150 MHz image, the XID+ model image, and the XID+ residual image. In the first column, the red plus signs indicate the positions of the LOFAR sources and the white dots indicate the positions of the IRAC sources (if present). The size of the symbol scales with the flux of the object. The SPIRE cutouts have a pixel size of 6\arcsec\,, and the LOFAR cutouts have a pixel size of 1.5\arcsec. All cutouts have a size of 54\arcsec $\times$ 54\arcsec.} 
\label{images}
\end{figure*}

Our de-blending approach for the bright {\it Herschel} 250 $\mu m$-selected sources is different from the default XID+ run for the LOFAR sources in two main ways. The first main difference is the input prior list. While we only considered LOFAR-detected sources, the default XID+ run for LOFAR uses IRAC sources as the input prior list, which has a much higher surface density. Figure~\ref{images} shows postage stamps of the original {\it Herschel} 250 $\mu$m image, the LOFAR 150-MHz image, the XID+ model image, and the XID+ residual image for a set of high-multiplicity {\it Herschel} sources (i.e. with three or more LOFAR counterparts within the {\it Herschel} beam in the Bo\"otes field).  In the first column where we show the original {\it Herschel} 250 $\mu$m image centred around the blind {\it Herschel} detection, we also overlay the positions of the LOFAR sources and the IRAC sources (if present). Comparing the {\it Herschel}  250 $\mu m$ images and the LOFAR 150-MHz images, it is clear that there is a high degree of correspondence between the FIR emission and the radio emission.

The second main difference is that we used flux priors based on the FIRC, while the default XID+ run for the LOFAR sources uses a flat flux prior between zero and the brightest value on the map. Previously, we found that using a flat flux prior can result in XID+ distributing the fluxes among the contributing sources in an overly democratic way. It is interesting to directly compare the SPIRE fluxes of our bright {\it Herschel} 250 $\mu m$-selected sources with the default SPIRE fluxes provided in the data release. As the default XID+ run for the LOFAR data release uses IRAC sources as an input list, our unique {\it Herschel}-LOFAR matches are also de-blended by the default XID+ run, while in this paper we adopted the fluxes from the blind {\it Herschel}-SPIRE source catalogues. The median of the flux ratio between the default XID+ flux and our adopted flux at 250 $\mu m$ for our unique match sample is 0.92 in ELAIS-N1. The median of the flux ratio between the default XID+ flux and our de-blended flux (using the SED prior enhanced XID+) at 250 $\mu m$ for our multiple match sample is 0.94 in ELAIS-N1.

\end{appendix}

\begin{acknowledgements}

{We thank the anonymous referee for providing constructive comments.}
We would like to thank the Center for Information Technology of the University of Groningen for their support and for providing access to the Peregrine high performance computing cluster. MJH acknowledges support from the UK Science and Technology Facilities Council (ST/R000905/1). KM acknowledges support from the National Science Centre (UMO-2018/30/E/ST9/00082). IM acknowledges support from STFC via grant ST/R505146/1. RK acknowledges support from the Science and Technology Facilities Council (STFC) through an STFC studentship via grant ST/R504737/1. MB acknowledges support from INAF under PRIN SKA/CTA FORECaST and from the Ministero degli Affari Esteri della Cooperazione Internazionale - Direzione Generale per la Promozione del Sistema Paese Progetto di Grande Rilevanza ZA18GR02. IP acknowledges support from INAF under the SKA/CTA PRIN "FORECaST" and PRIN MAIN STREAM "SAuROS" projects. PNB and JS are grateful for support from the UK STFC via grant ST/R000972/1. MV acknowledges support from the Italian Ministry of Foreign Affairs and International Cooperation (MAECI Grant Number ZA18GR02) and the South African Department of Science and Technology's National Research Foundation (DST-NRF Grant Number 113121) as part of the ISARP RADIOSKY2020 Joint Research Scheme.

The Herschel Extragalactic Legacy Project, (HELP), is a European Commission Research Executive Agency funded project under the SP1-Cooperation, Collaborative project, Small or medium-scale focused research project, FP7-SPACE-2013-1 scheme, Grant Agreement Number 607254.

LOFAR is the Low Frequency Array designed and constructed by ASTRON. It has observing, data processing, and data storage facilities in several countries, which are owned by various parties (each with their own funding sources), and which are collectively operated by the ILT foundation under a joint scientific policy. The ILT resources have benefitted from the following recent major funding sources: CNRS-INSU, Observatoire de Paris and Universit\'e d'Orl\'eans, France; BMBF, MIWF-NRW, MPG, Germany; Science Foundation Ireland (SFI), Department of Business, Enterprise and Innovation (DBEI), Ireland; NWO, The Netherlands; The Science and Technology Facilities Council, UK; Ministry of Science and Higher Education, Poland. This research made use of the Dutch national e-infrastructure with support of the SURF Cooperative (e-infra 180169) and the LOFAR e-infra group. The J\"ulich LOFAR Long Term Archive and the German LOFAR network are both coordinated and operated by the J\"ulich Supercomputing Centre (JSC), and computing resources on the Supercomputer JUWELS at JSC were provided by the Gauss Centre for Supercomputing e.V. (grant CHTB00) through the John von Neumann Institute for Computing (NIC). This research made use of the University of Hertfordshire high-performance computing facility and the LOFAR-UK computing facility located at the University of Hertfordshire and supported by STFC [ST/P000096/1].

\end{acknowledgements}

\end{document}